\documentclass[aps,prd,12pt,preprint,superscriptaddress,preprintnumbers,nofootinbib,floatfix,longbibliography]{revtex4-1}

\usepackage[T1]{fontenc}
\usepackage[utf8]{inputenc}
\usepackage[english]{babel}
\usepackage{amsmath,amsthm,amssymb,amsfonts,mathrsfs,amsbsy,bm}
\usepackage{tensor}
\usepackage{slashed}
 \usepackage{mathrsfs,tikz} 
\usepackage{esint}
\usepackage[a4paper, margin=1.5cm]{geometry}
\usepackage{cancel}
\usepackage{placeins}
\usepackage{graphicx}
\usepackage{multirow}
\usepackage{array}
\usepackage{booktabs}
\usepackage{makecell}
\usepackage{xcolor}
\colorlet{BLUE}{blue}
\usepackage{tikz}
\usetikzlibrary{quotes,angles,arrows,decorations.markings,decorations.pathmorphing}
\usepackage{hyperref}
\hypersetup{colorlinks=true,breaklinks=true,citecolor=blue,linkcolor=[rgb]{0,0.5,0.9},urlcolor=blue}
\graphicspath{{figures/}}

\newcommand{\be}{\begin{equation}}
\newcommand{\ee}{\end{equation}}
\newcommand{\Be}{\begin{eqnarray}}
\newcommand{\Ee}{\end{eqnarray}}

\def\e{\mathrm{e}}

\newcommand{\dd}{\mathrm{d}}
\DeclareMathOperator{\Tr}{Tr}
\newcommand{\ket}[1]{\lvert#1\rangle}
\newcommand{\bra}[1]{\langle#1\rvert}

\newcommand{\mincir}{\raise-3.truept\hbox{\rlap{\hbox{$\sim$}}\raise4.truept\hbox{$<$}\ }}
\newcommand{\magcir}{\raise-3.truept\hbox{\rlap{\hbox{$\sim$}}\raise4.truept\hbox{$>$}\ }}

\providecommand{\U}[1]{}

\newcommand{\ie}{\begin{equation}}
\newcommand{\fe}{\end{equation}}
\newcommand{\se}{\begin{eqnarray}}
\newcommand{\ff}{\end{eqnarray}}

\providecommand{\preparepaperbibliography}{}

\begin{document}
\emergencystretch=4em
\usetikzlibrary{arrows.meta,patterns}


\title{Entanglement and HBAR entropy in a Lorentzian non--commutative black hole}

\author{A. A. Ara\'{u}jo Filho}
\email{dilto@fisica.ufc.br}
\affiliation{Departamento de F\'isica, Universidade Federal da Para\'iba, Caixa Postal 5008, 58051--970, Jo\~ao Pessoa, Para\'iba, Brazil.}
\affiliation{Departamento de F\'isica, Universidade Federal de Campina Grande, Caixa Postal 10071, 58429--900 Campina Grande, Para\'iba, Brazil.}
\affiliation{Center for Theoretical Physics, Khazar University, 41 Mehseti Street, Baku, AZ-1096, Azerbaijan.}

\author{N.~Heidari}
\email{heidari.n@gmail.com}

\affiliation{Departamento de F\'isica, Universidade Federal de Campina Grande, Caixa Postal 10071, 58429--900 Campina Grande, Para\'iba, Brazil.}
\affiliation{Center for Theoretical Physics, Khazar University, 41 Mehseti Street, Baku, AZ-1096, Azerbaijan.}

\author{P. H. M. Barros}
\email{phmbarros@ufpi.edu.br}
\affiliation{Departamento de Matemática e F\'{i}sica, Universidade Estadual do Maranhão (UEMA), Caxias - MA, 65604-380, Brazil}

\author{V. B. Bezerra}
\email{valdir@fisica.ufpb.br}
\affiliation{Departamento de F\'isica, Universidade Federal da Para\'iba, Caixa Postal 5008, 58051--970, Jo\~ao Pessoa, Para\'iba, Brazil.}

\author{A. R. Queiroz}
\email{amilcarq@df.ufcg.edu.br}
\affiliation{Departamento de F\'isica, Universidade Federal de Campina Grande, Caixa Postal 10071, 58429--900, Campina Grande, Para\'iba, Brazil.}

\author{A. F. Santos}
\email{alesandroferreira@fisica.ufmt.br}
\affiliation{Programa de Pós-graduação em Física, Instituto de Física, Universidade Federal de Mato Grosso, Cuiabá, Brasil}

\author{Ali \"Ovg\"un}
\email{ali.ovgun@emu.edu.tr}
\affiliation{
Physics Department, Eastern Mediterranean University, Famagusta, 99628 North Cyprus via Mersin 10, Turkiye.}


\date{\today}


\let\paperbodyspread\baselinestretch
\linespread{1.0}\selectfont
\begin{abstract}

We investigate how Lorentzian non--commutative smearing affects entanglement and horizon brightened acceleration radiation (HBAR) outside a black hole. For a massless scalar field, tracing over the interior partner gives the exterior quantum channel, its logarithmic negativity, mutual information, and Bell entanglement fidelity. At fixed mass and Killing frequency, increasing the smearing lowers the surface gravity, increases the accessible correlations, and suppresses the leading atomic emission probability. The horizon displacement begins at order $\sqrt{\Theta}$, while the thermal correction at infinity begins at order $\Theta$. At fixed local frequency, the redshift introduces a term of order $\sqrt{\Theta}$ at fixed areal radius. Holding the proper distance fixed, it cancels this term and recovers the common Rindler response close to the horizon. For a freely falling atom, emission and absorption obey detailed balance at the Hawking temperature in the leading horizon approximation. Random atomic injection drives the selected mode from the vacuum to a thermal state with an exact relaxation law and nonnegative entropy production. Near equilibrium, we obtain $\dot S_{\rm HBAR}\simeq\dot A_p/4$ when the radiated energy is drawn from horizon heat, where $\dot A_p$ is the positive radiative area loss. When the radiation is assigned to ADM mass loss at fixed smearing length $\ell=4\sqrt{\Theta/\pi}$, the result is $\dot S_{\rm HBAR}\simeq\dot A_p/[4(1-\ell/r_+)]$. The mass dependence of the exterior source thus enters the entropy per unit area lost at order $\sqrt{\Theta}$.

\end{abstract}

\maketitle
\linespread{\paperbodyspread}\selectfont



\section{Introduction}
\label{sec:introduction}

Hawking's calculation relates the temperature of a black hole to the surface gravity of its horizon, $T_H=\kappa/(2\pi)$~\cite{Hawking1975,HartleHawking1976}. For an outgoing scalar mode in the vacuum regular across the future horizon, the ratio $\omega/T_H$ fixes the mean exterior occupation and the mixing with an interior partner~\cite{Unruh1976,Israel1976}. A static observer measures the frequency and the temperature with the same gravitational redshift~\cite{Tolman1930}. Across a family of black holes, however, fixing the frequency at infinity selects a different comparison from fixing the frequency on the observer's clock. If the horizon itself moves as a parameter varies, an observer held at the same areal radius also samples the field at a different proper distance from it.

The idea that spacetime admits a finite resolution goes back to Snyder, who represented its coordinates by noncommuting operators compatible with Lorentz invariance~\cite{Snyder1947}. Doplicher, Fredenhagen, and Roberts related localization bounds to the gravitational effect of the energy needed to resolve an event~\cite{Doplicher1995}, while Seiberg and Witten obtained non--commutative gauge theories in an open string limit with an antisymmetric background field~\cite{SeibergWitten1999}. In gauge formulations of gravity, Chaichian, Tureanu, and Zet applied the Seiberg--Witten map to the Schwarzschild geometry~\cite{Chaichian2008Schwarzschild}. More recently, Juri\'c, Naveena Kumara, and Po\v{z}ar showed how the choice of Moyal twist affects the horizon geometry~\cite{Juric2025BlackHoles}. Thermodynamic studies of deformed Schwarzschild metrics have examined evaporation~\cite{AraujoFilho2023GaugeThermo} and, for a twist coupling time to the polar angle, the tunneling probabilities of bosons and fermions~\cite{AraujoFilho2026Gauge}. A perturbation of the metric coefficients also changes the effective potential for scalar scattering and the corresponding quasinormal frequencies~\cite{Heidari2024Perturbation,Heidari:2025iiv,AraujoFilho:2025jcu}. A different prescription places the deformation in the Schwarzschild mass parameter. It has been used to study scalar absorption and the shadow~\cite{Heidari2024Stable}, as well as light travel times and the phases of massive neutrinos~\cite{AraujoFilho2025Geodesics}.

In the description adopted here, the length scale enters through the matter distribution in the Einstein equations. The Gaussian source of Nicolini, Smailagic, and Spallucci gives a regular de Sitter core and approaches the Schwarzschild geometry at large radii~\cite{Nicolini2006}. Nozari and Mehdipour subsequently studied Hawking emission by tunneling in this geometry~\cite{Nozari2008Tunneling}. A Lorentzian source has an algebraic tail, which contributes a term proportional to $M\sqrt{\Theta}/r^2$ to the exterior lapse where $M$ is the mass at infinity and $\Theta$ has dimensions of length squared. This term affects scalar absorption and scattering~\cite{Anacleto2020Scattering}, quasinormal frequencies and the shadow~\cite{Campos2022}, and gravitational deflection~\cite{AraujoFilho2024Lorentzian}. Its rotating extension also allows the dependence of the shadow and scalar spectrum on angular momentum to be examined~\cite{AraujoFilho2025Axisymmetric}.

The radial dependence of Lorentzian smearing can be followed along individual trajectories. It changes the phase acquired by massive neutrinos~\cite{AraujoFilho2026Neutrino} and the intensity received from a thin emitting disk~\cite{AraujoFilho2026Disk}. Moreover, Heidari \textit{et al.} found that increasing the smearing scale moves the innermost stable circular orbit inward and shifts the phase of the gravitational waves calculated from periodic orbits~\cite{Heidari2026Periodic}. Particle creation probes a different feature of the same metric. At fixed $M$, the outer horizon shifts at order $\sqrt{\Theta}$, yet the term of this order cancels in the Hawking temperature. The first thermal correction at infinity is of order $\Theta$~\cite{AraujoFilho2025Particles}. Which of these orders governs the entanglement accessible outside the horizon, and which enters the radiation emitted by a freely falling atom?

Fuentes-Schuller and Mann showed how acceleration reduces the entanglement accessible to observers sharing a bosonic state~\cite{FuentesMann2005}. Pan and Jing related this loss to the Hawking temperature in static black hole spacetimes~\cite{PanJing2008}. For a scalar field prepared in a two mode squeezed state, Adesso and Fuentes-Schuller showed that the loss of exterior bipartite entanglement accompanies its redistribution among modes on both sides of the horizon~\cite{Adesso2009Horizon}. Near a Schwarzschild horizon, the accessible correlations depend on the observer's position, the field frequency, and the choice of wave packet~\cite{MartinMartinez2010,Bruschi2010}. Propagation to infinity introduces a further dependence on the exterior scattering potential, as Br\'adler and Adami showed in their analysis of bosonic quantum channels~\cite{BradlerAdami2015}. More recently, Sen, Mukherjee, and Gangopadhyay calculated logarithmic negativity, mutual information, and entanglement fidelity for a quantum corrected Schwarzschild geometry~\cite{Sen2024Entanglement}. Related studies have compared regular and singular black holes and examined the influence of Lorentz violation and effective quantum corrections on accessible entanglement~\cite{Luongo2026Entanglement,Liu2025LorentzViolation,Liu2026Effective}. In non--commutative models, previous work has followed the spin entanglement of two propagating particles~\cite{Mohadi2022Spin} and the response of a static atomic detector outside a black hole with a Gaussian source~\cite{Shi2024Detector}.

A freely falling atom provides a complementary probe. Scully \textit{et al.} showed that an atom initially in its ground state can become excited while emitting a quantum into an exterior vacuum mode. A stream of such atoms produces horizon brightened acceleration radiation (HBAR)~\cite{Scully2018}. The outgoing phase along the trajectory accounts for the thermal factor~\cite{Camblong2020,AziziKerr2021}, and random atomic injection relates the radiation entropy to the radiative change in horizon area~\cite{AziziMaster2021,AziziThermodynamics2021}. This relation has been explored for quantum corrected and regular black holes~\cite{Sen2022HBAR,Jana2025HBAR,Uktamov2026HBAR} and used to bound the information carried by the radiation~\cite{Maleki2026}. In bumblebee gravity, Ara\'ujo Filho and Liu showed that vacuum configurations with identical exterior geometries give the same scalar correlations, although their conserved energies can lead to different coefficients in the relation between entropy and area~\cite{AraujoLiu2025}. Lorentzian smearing raises a related question within Einstein gravity itself, since varying $M$ also changes the energy stored in the extended matter source~\cite{MaZhao2014}.

In this work, we entangle a reference mode with a narrow outgoing packet of a massless scalar field and derive the exterior quantum channel by tracing over the interior partner. At fixed $M$ and Killing frequency $\omega$, increasing $\Theta$ lowers $\kappa$ and increases the logarithmic negativity, mutual information, and Bell entanglement fidelity. For a freely falling atom coupled to a packet prepared in the exterior Boulware vacuum, the same factor $e^{-2\pi\omega/\kappa}$ fixes the leading emission to absorption ratio. At fixed atomic gap and common packet normalization, the leading emission probability decreases. This thermal contribution and the scalar correlations first change at order $\Theta$ for fixed $\omega$. A finite interaction also samples the radial profile of the selected mode. For the scalar correlations, fixing the local frequency and areal radius admits an earlier contribution of order $\sqrt{\Theta}$ through the lapse at the observer. Moreover, at fixed local frequency $\omega_{\mathrm{loc}}$ and small proper distance $\varrho$, the thermal factor reduces to $e^{-2\pi\omega_{\mathrm{loc}}\varrho}$ at leading order. We show that the cancellation of the term linear in the smearing length extends to every finite proper distance in the effective geometry.

Finally, random atomic injection gives an exact relaxation law from the vacuum to a thermal stationary state. Near equilibrium, the field entropy satisfies $\dd S_{\mathrm{HBAR}}\simeq\dd E_{\mathrm{rad}}/T_H$. At fixed smearing length $\ell=4\sqrt{\Theta/\pi}$, the effective exterior obeys $T_H\,\dd(A/4)=(1-\ell/r_+)\,\dd M$, where $r_+$ is the outer horizon radius. Wang \textit{et al.} obtained the same leading source factor in the equilibrium first law for a Lorentzian black hole in anti-de Sitter spacetime~\cite{Wang2025LorentzianFirstLaw}. Here, the factor enters when the incremental packet energy is assigned to a particular radiative change of the black hole. Identifying the radiated energy with horizon heat gives the usual one quarter coefficient, while identifying it with the loss of mass at infinity gives $1/[4(1-\ell/r_+)]$. The matter outside the horizon thus contributes at order $\sqrt{\Theta}$ to the relation between radiative entropy and area, even though the temperature at infinity first changes at order $\Theta$.


\section{Lorentzian smearing and the exterior geometry}\label{sec:geometry}
\subsection{Density, enclosed mass, and the adopted lapse}
We adopt the Lorentzian non--commutative distribution used in Ref.~\cite{AraujoFilho2024Lorentzian}, with the smearing placed in the matter source of the Einstein equations. In units $G=c=\hbar=k_B=1$, the density and its enclosed mass are
\begin{equation} 
\begin{aligned} 
\rho_\Theta(r)&=\frac{M\sqrt{\Theta}}{\pi^{3/2}(r^2+\pi\Theta)^2},\qquad a_\Theta=\sqrt{\pi\Theta},\\ m_\Theta(r)&=4\pi\int_0^r u^2\rho_\Theta(u)\,\dd u=\frac{2M}{\pi}\left[\arctan\!\left(\frac{r}{a_\Theta}\right)-\frac{a_\Theta r}{r^2+a_\Theta^2}\right],
\end{aligned} \label{eq:lorentzianmass}
\end{equation}
where $M$ is the mass measured at infinity and $\Theta$ has dimension of length squared. The normalization follows from $m_\Theta(\infty) = M$. Lorentzian and Gaussian smearing both replace a point source by a smooth distribution, with distinct radial tails and thermodynamic scales \cite{Nicolini2006,Anacleto2020Scattering,Anacleto2021Entropy}. Introducing the length $\ell=4\sqrt{\Theta/\pi}=4a_\Theta/\pi$, the large radius expansion of Eq.~\eqref{eq:lorentzianmass} gives
\begin{equation} 
\frac{m_\Theta(r)}{M}=1-\frac{\ell}{r}+\frac{\pi^2\ell^3}{24r^3}+O\!\left(\frac{\ell^5}{r^5}\right),\qquad f_{\rm full}(r)=1-\frac{2m_\Theta(r)}r=1-\frac{2M}{r}+\frac{2M\ell}{r^2}-\frac{\pi^2M\ell^3}{12r^4}+O\!\left(\frac{M\ell^5}{r^6}\right). 
\label{eq:fullseries}
\end{equation}
The exterior model studied below is the leading Lorentzian lapse of Ref.~\cite{AraujoFilho2024Lorentzian},
\begin{equation} 
\dd s^2=-f(r)\dd t^2+\frac{\dd r^2}{f(r)}+r^2\dd\Omega_2^2,\qquad f(r)=1-\frac{2M}{r}+\frac{2M\ell}{r^2}=1-\frac{2M}{r}+\frac{8M\sqrt{\Theta}}{\sqrt{\pi}\,r^2}. 
\label{eq:metric}
\end{equation}
The condition $f(\infty)=1$ fixes the normalization of $t$. Equation~\eqref{eq:metric} defines the effective geometry throughout the quantum calculation. The full density in Eq.~\eqref{eq:lorentzianmass} supplies an additional control of the smearing expansion. In particular, the first omitted lapse term relative to its leading smearing correction has magnitude $\pi^2\ell^2/(24r^2)$. 

The effective enclosed mass is $m(r)=M-M\ell/r$. For the reciprocal lapse metric, the Einstein equations give $\rho=m'/(4\pi r^2)$, $p_r=-\rho$, and $p_t=-m''/(8\pi r)$. Their evaluation yields
\begin{equation}
\rho_{\rm eff}(r)=\frac{M\ell}{4\pi r^4},\qquad p_r=-\rho_{\rm eff},\qquad p_t=\rho_{\rm eff}. 
\label{eq:effectivesource}
\end{equation}
This source is the exterior expansion of the Lorentzian distribution. Its explicit mass dependence enters the horizon energy balance in Sec.~\ref{sec:hbar_area}. The radial function shares the algebraic form of a Reissner--Nordstr\"om lapse, with coefficient $C=2M\ell$ fixed by the smearing prescription. This correspondence permits direct comparisons with charged horizon geometries while keeping the source specified by Eq.~\eqref{eq:effectivesource} \cite{Anacleto2020Scattering,Luongo2026Entanglement}.
\subsection{Horizons, surface gravity, and proper distance}

A simple outer horizon exists for $M>2\ell$. For $0<\ell<M/2$, the two Killing horizons, their separation, and the factorized lapse are
\begin{equation} 
r_\pm=M\pm\sqrt{M^2-2M\ell},\qquad \Delta_h=r_+-r_-,\qquad f(r)=\frac{(r-r_+)(r-r_-)}{r^2}. 
\label{eq:horizons}
\end{equation}
We use $\Delta_h$ for the horizon separation and $\dd\Omega_2^2$ for the unit sphere metric. At $\ell=0$, the outer root is the Schwarzschild horizon and the inner root parameter reaches the central singularity. The derivative of the factorized lapse at $r_+$ gives 
\begin{equation} \kappa=\frac{f'(r_+)}2=\frac{\Delta_h}{2r_+^2},\qquad T_H=\frac{\kappa}{2\pi}=\frac{\Delta_h}{4\pi r_+^2}. 
\label{eq:kappa}
\end{equation}
Their role in particle creation follows from the regularity of horizon crossing modes \cite{Hawking1975,HartleHawking1976,Unruh1976}. For a static observer at $r_o>r_+$, the four velocity is $u^\mu=f(r_o)^{-1/2}\delta^\mu_t$. The radial acceleration and its magnitude are $a^r=f'(r_o)/2$ and $a_{\rm prop}=|f'(r_o)|/[2\sqrt{f(r_o)}]$. Tolman redshift gives
\begin{equation}
\omega_{\rm loc} = \frac{\omega}{\sqrt{f(r_o)}},\qquad T_{\rm loc}=\frac{T_H}{\sqrt{f(r_o)}},\qquad a_R=\frac{\kappa}{\sqrt{f(r_o)}},\qquad \frac{a_{\rm prop}}{a_R}=\frac{|f'(r_o)|}{2\kappa},
\label{eq:tolman}
\end{equation}
where $\omega$ is the frequency conjugate to the normalized time $t$ \cite{Tolman1930}. The quantity $a_R=2\pi T_{\rm loc}$ characterizes the thermal scale, while $a_{\rm prop}$ describes the actual static worldline. Their ratio approaches unity at the outer horizon.

Set $x = r - r_{+}$. The proper radial distance from the horizon on a static slice has the closed form
\begin{equation} 
\varrho(r)=\int_{r_+}^{r}\frac{\dd u}{\sqrt{f(u)}}=\sqrt{x(x+\Delta_h)}+2M\,\operatorname{arsinh}\sqrt{\frac{x}{\Delta_h}}. 
\label{eq:properdistance}
\end{equation}
For $x\ll\Delta_h,r_+$, the local geometry becomes
\begin{equation} 
x = \frac{\kappa\varrho^2}{2}+\frac{\kappa f''_+\varrho^4}{48}+O(\varrho^6),\qquad \sqrt{f(r)}=\kappa\varrho\left[1+\frac{f''_+\varrho^2}{12}+O(\varrho^4)\right],\qquad f''_+=\frac{2(2r_--r_+)}{r_+^3}. 
\label{eq:rindlerexpansion}
\end{equation}
In this manner, $\dd s^2\simeq-\kappa^2\varrho^2\dd t^2+\dd\varrho^2+r_+^2\dd\Omega_2^2$. With $T=\varrho\sinh(\kappa t)$ and $X=\varrho\cosh(\kappa t)$, the radial part is locally Minkowskian. The first correction in Eq.~\eqref{eq:rindlerexpansion} also quantifies the geometric contribution to a comparison made at fixed proper distance. This local construction connects the scalar correlation analysis to the Schwarzschild calculation of Ref.~\cite{MartinMartinez2010} and to recent comparisons among charged and regular geometries \cite{Luongo2026Entanglement}.

The distinction between the effective and complete smeared geometries becomes especially useful near the limiting mass. Equation~\eqref{eq:horizons} gives the effective double root $M_e = r_e  = 2\ell = 8\sqrt{\Theta/\pi}$. For the full distribution, the simultaneous conditions $f_{\rm full}=f'_{\rm full}=0$ instead give $r_e/a_\Theta=1.82525564252$ and $M_e/a_\Theta=2.21157049140$. The first omitted term in Eq.~\eqref{eq:fullseries} reaches about $10.3\%$ of the leading smearing correction at the effective double root. We interpret the approach to $\Delta_h=0$ as a limit of the adopted effective geometry, and use the full source to quantify its relation to Lorentzian smearing.
\subsection{Two scales in the small smearing expansion}

At fixed mass, write $\eta=\ell/M$ and $s_h=\sqrt{1-2\eta}$. Equation~\eqref{eq:kappa} gives
\begin{equation}
4M\kappa=\frac{4s_h}{(1+s_h)^2}=1-\left(\frac{1-s_h}{1+s_h}\right)^2,\qquad \frac{\partial\kappa}{\partial\eta}=-\frac{1-s_h}{M s_h(1+s_h)^3}. 
\label{eq:kappamonotone}
\end{equation}
The surface gravity decreases as the smearing length increases. Expanding $s_h$ about the Schwarzschild value gives
\begin{equation} \begin{aligned}r_+&=2M-\ell-\frac{\ell^2}{2M}+O\!\left(\frac{\ell^3}{M^2}\right),\\4M\kappa&=1-\frac{\ell^2}{4M^2}-\frac{\ell^3}{2M^3}+O\!\left(\frac{\ell^4}{M^4}\right)=1-\frac{4\Theta}{\pi M^2}+O\!\left(\frac{\Theta^{3/2}}{M^3}\right).\end{aligned} \label{eq:smalltheta}\end{equation}
A horizon displacement of order $\sqrt{\Theta}$ accompanies an asymptotic thermal correction of order $\Theta$. The cancellation at linear order in $\ell$ will organize both the entanglement and HBAR results. For the full source, the expansion $4M\kappa_{\rm full}=1-\eta^2/4+(-1/2+\pi^2/96)\eta^3+O(\eta^4)$ shares this leading thermal correction of order $\Theta$. At fixed $\ell$, the temperature may also be written as $T_H=(r_+-2\ell)/[4\pi r_+(r_+-\ell)]$. It attains its maximum at $r_+=(2+\sqrt2)\ell$, or $M=(1+\sqrt2)\ell$, with $T_{\max}=(3-2\sqrt2)/(4\pi\ell)$. 
\section{Scalar modes and the horizon transformation}\label{sec:modes}
\subsection{Radial equation and horizon crossing modes}

Consider a real, massless, minimally coupled scalar field satisfying $\Box\Phi=0$. Separation with $\Phi_{\omega jm}=r^{-1}\psi_{\omega j}(r)Y_{jm}(\theta,\phi)\e^{-i\omega t}$ yields
\begin{equation}
\frac{\dd^2\psi_{\omega j}}{\dd r_*^2}+\left[\omega^2-V_j(r)\right]\psi_{\omega j}=0,\qquad V_j=f(r)\left[\frac{j(j+1)}{r^2}+\frac{f'(r)}r\right],\qquad \frac{\dd r_*}{\dd r}=\frac1f.
\label{eq:KG}
\end{equation}
The potential vanishes at the simple horizon, where the radial equation admits ingoing and outgoing plane waves. Integrating $\dd r_*/\dd r=1/f$ for the effective lapse gives
\begin{equation}
r_*=r+\frac{r_+^2}{\Delta_h}\ln\!\left(\frac{r-r_+}{L_0}\right)-\frac{r_-^2}{\Delta_h}\ln\!\left(\frac{r-r_-}{L_0}\right),
\label{eq:tortoise}
\end{equation}
where $L_0>0$ fixes an irrelevant additive constant and $r>r_+$. Using Eq.~\eqref{eq:tortoise}, define $u=t-r_*$ and $v=t+r_*$. Outgoing static modes have the form $\e^{-i\omega u}$. The coordinates $U=-\kappa^{-1}\e^{-\kappa u}$ and $V=\kappa^{-1}\e^{\kappa v}$ extend the local description through the horizon \cite{Hawking1975,Boulware1975,HartleHawking1976}. Outside it, the outgoing mode is proportional to $(-\kappa U)^{i\omega/\kappa}$. Continuing through the lower half of the complex $U$ plane selects positive freely falling frequency. For $U>0$, the logarithm acquires $+i\pi$, giving the interior piece an amplitude $\e^{-\pi\omega/\kappa}$ relative to the exterior piece.

The scalar product $(\phi_1,\phi_2)=i\int_\Sigma\dd\Sigma^\mu(\phi_1^*\nabla_\mu\phi_2-\phi_2\nabla_\mu\phi_1^*)$ fixes the normalization. For one selected outgoing pair, the regular annihilation operator and its thermal parameter are
\begin{equation}
b_H=\frac{b_B-\sqrt q\,b_C^\dagger}{\sqrt{1-q}},\qquad q=\e^{-2\pi\omega/\kappa}=\e^{-\omega/T_H},\qquad |\alpha_\omega|^2=\frac1{1-q},\qquad |\beta_\omega|^2=\frac q{1-q}.
\label{eq:bogoliubov}
\end{equation}
The labels $B$ and $C$ denote the accessible exterior oscillator and its interior partner. Equation~\eqref{eq:bogoliubov} implements the horizon mixing underlying the Hawking and Unruh effects \cite{Fulling1973,Unruh1976}. We use outgoing wave packets centered at $\omega$, with width $\Delta\omega\ll\omega$. Their squeezing parameter $\zeta=\operatorname{arctanh}\sqrt q$ satisfies $|\partial_\omega\zeta|=\pi\sqrt q/[\kappa(1-q)]$, so $\pi\sqrt q\,\Delta\omega/[\kappa(1-q)]\ll1$ makes the squeezing nearly constant across each packet. These conditions specify the mode resource and follow the wave packet interpretation discussed in Ref.~\cite{Bruschi2010}.

We choose the horizon analytic outgoing state regular at the future horizon. The Unruh state supplies this sector for an evaporating black hole, whereas the Hartle--Hawking state also includes an incoming thermal sector. Static exterior oscillators are defined with the Boulware frequency splitting \cite{Boulware1975,HartleHawking1976,Unruh1976}. The correlations below describe the ideal outgoing horizon resource in the conformal radial sector. The local frequency curves select this resource through the Tolman relation. Spatially resolved detection and propagation to distant observers additionally use the radial mode functions and potential in Eq.~\eqref{eq:KG}.
\subsection{Vacuum and one particle isometry}

Write $d=1-q$. The regular vacuum of the selected pair is annihilated by $b_H$ and by the complementary operator $\widetilde b_H=(b_C-\sqrt q\,b_B^\dagger)/\sqrt d$. The two mode transformation preserves the number difference, $b_H^\dagger b_H-\widetilde b_H^\dagger\widetilde b_H=b_B^\dagger b_B-b_C^\dagger b_C$. The vacuum therefore contains equal occupations in $B$ and $C$, so $\ket0_H=\sum_{n\ge0}c_n\ket n_B\ket n_C$. Acting with $b_H$ gives $c_{n+1}=\sqrt q\,c_n$, and normalization fixes $c_0=\sqrt d$ up to an overall phase. Applying $b_H^\dagger$ then gives the one particle state. The two states are
\begin{equation}
\ket0_H=\sqrt d\sum_{n=0}^\infty q^{n/2}\ket n_B\ket n_C,\qquad \ket1_H=d\sum_{n=0}^\infty\sqrt{n+1}\,q^{n/2}\ket{n+1}_B\ket n_C.
\label{eq:isometry}
\end{equation}
The geometric identities $\sum_nq^n=d^{-1}$ and $\sum_n(n+1)q^n=d^{-2}$ verify both norms. Tracing the interior vacuum partner gives $\rho_{B,0}=d\sum_nq^n\ket n\bra n$ and $\langle n\rangle=q/d$. The same parameter fixes the thermal population and the correlations shared across the horizon \cite{FuentesMann2005,MartinMartinez2010,Sen2024Entanglement,AraujoLiu2025}.
\section{Accessible entanglement, mutual information, and the scalar channel}\label{sec:correlations}
\subsection{Bell resource and the reduced density operator}

\begin{figure}[!htbp]\centering\includegraphics[width=0.7\textwidth]{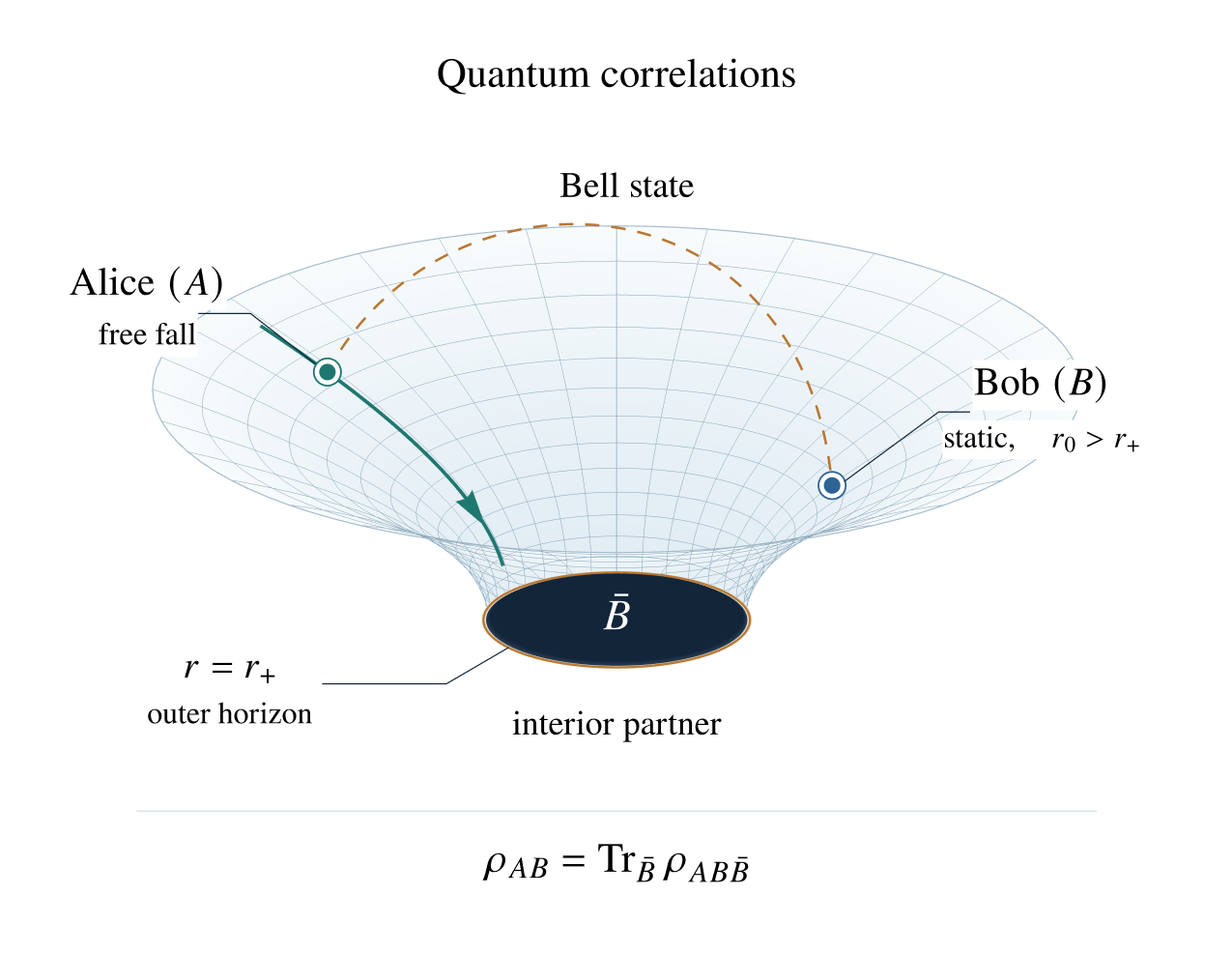}\caption{Entanglement setup in the Lorentzian non--commutative exterior. Alice ($A$) follows radial free fall, Bob ($B$) remains static at $r_0=r_o>r_+$, and $\bar B\equiv C$ labels the inaccessible interior partner. The dashed curve represents the initial Bell correlation. The mesh illustrates an auxiliary embedding of the static equatorial exterior, with the dark region marking the schematic interior.}\label{fig:lorentzian_entanglement_setup}\end{figure}

Alice holds a reference qubit $A$ initially entangled with the selected regular mode $H$. The transformation in Eq.~\eqref{eq:isometry} acts only on $H$, while $A$ serves as a reference for comparing the initial and exterior correlations. The initial and final pure states are
\begin{equation}
\ket\Psi_{AH}=\frac{\ket0_A\ket0_H+\ket1_A\ket1_H}{\sqrt2},\qquad \ket\Psi_{ABC}=\sum_{n=0}^\infty\sqrt{\frac{dq^n}{2}}\left[\ket{0,n}_{{AB}}+\sqrt{d(n+1)}\ket{1,n+1}_{{AB}}\right]\ket n_C.
\label{eq:Bell}
\end{equation}
Figure~\ref{fig:lorentzian_entanglement_setup} places Alice in free fall and Bob at a fixed exterior radius. The oscillators $B$ and $\bar B\equiv C$ are the exterior and interior outputs associated with $H$.
The orthonormal interior basis in Eq.~\eqref{eq:Bell} makes its trace select equal values of $n$ and gives
\begin{equation} \begin{aligned}\rho_{AB}=\frac d2\sum_{n=0}^\infty q^n\Big\{&\ket{0,n}\bra{0,n}+d(n+1)\ket{1,n+1}\bra{1,n+1}\\&+\sqrt{d(n+1)}\big[\ket{0,n}\bra{1,n+1}+\ket{1,n+1}\bra{0,n}\big]\Big\}.\end{aligned} \label{eq:rhoAB}\end{equation}
Each term of Eq.~\eqref{eq:rhoAB} occupies the two dimensional subspace spanned by $\ket{0,n}$ and $\ket{1,n+1}$. Its matrix is $(dq^n/2)\bigl(\begin{smallmatrix}1&\sqrt{d(n+1)}\\ \sqrt{d(n+1)}&d(n+1)\end{smallmatrix}\bigr)$, with one positive eigenvalue and one zero eigenvalue. The joint eigenvalues and the two marginals are
\begin{equation}
p_n=\frac d2q^n[1+d(n+1)],\qquad \rho_A=\frac{\mathbb I_2}{2},\qquad \rho_B=\sum_{m=0}^\infty b_m\ket m\bra m,\qquad b_m=\frac d2q^m+\frac{d^2}{2}m q^{m-1}.
\label{eq:spectra}
\end{equation}
The second term in $b_m$ is defined as zero at $m=0$. Summing the geometric series verifies $\sum_np_n=\sum_mb_m=1$.
\subsection{Partial transpose and logarithmic negativity}

We use $\mathcal N=(\|\rho_{AB}^{T_A}\|_1-1)/2$ and $E_{\mathcal N}=\log_2(1+2\mathcal N)$ \cite{VidalWerner2002}. Partial transposition sends $\ket{0,n}\bra{1,n+1}$ to $\ket{1,n}\bra{0,n+1}$. It leaves an isolated positive entry $d/2$ and gives the following blocks in the basis $\{\ket{0,n+1},\ket{1,n}\}$,
\begin{equation} \begin{aligned} B_n&=\begin{pmatrix}A_n&C_n\\ C_n&D_n\end{pmatrix},\qquad \det B_n=-\frac{d^3q^{2n}}4,\\ A_n&=\frac d2q^{n+1},\qquad D_n=\frac{d^2}{2}nq^{n-1},\qquad C_n=\frac{d^{3/2}}2\sqrt{n+1}\,q^n.\end{aligned} \label{eq:PTblocks}\end{equation}
Here, $D_0=0$. The two eigenvalues are $\lambda_n^\pm=[A_n+D_n\pm\sqrt{(A_n-D_n)^2+4C_n^2}]/2$. For $0<q<1$, the negative determinant gives one negative eigenvalue per block. Rationalizing its magnitude gives the positive series
\begin{equation} \begin{aligned}\mathcal N&=\sum_{n=0}^\infty\mathcal N_n,\qquad \mathcal N_0=\frac{d^2}{2},\qquad E_{\mathcal N}=\log_2(1+2\mathcal N),\\ \mathcal N_n&=\frac{d^2q^{n+1}}{\sqrt{(q^2+nd)^2+4dq^2}+q^2+nd}\quad(n\ge1).\end{aligned} \label{eq:negativity}\end{equation}
The zero thermal occupation limit $q\to0$ gives $\mathcal N=1/2$ and $E_{\mathcal N}=1$, while $q\to1$ gives $\mathcal N\to0$. These limits show how the interior trace changes the entanglement accessible to Alice and Bob \cite{FuentesMann2005,MartinMartinez2010}. Equation~\eqref{eq:negativity} also provides a direct comparison with quantum corrected and effective geometries studied in Refs.~\cite{Sen2024Entanglement,Liu2025LorentzViolation,Liu2026Effective,Luongo2026Entanglement}.
\subsection{Mutual information and certified sums}

With logarithms in base two and $0\log_2 0=0$, define $H(\{z_n\})=-\sum_nz_n\log_2z_n$. Equation~\eqref{eq:spectra} gives
\begin{equation}
\begin{aligned} S_{AB}&=H(\{p_n\}),\qquad S_B=H(\{b_n\}),\qquad S(A|B)=S_{AB}-S_B,\\ I(A:B)&=1+S_B-S_{AB}=1+\sum_{n=0}^{\infty}\big[p_n\log_2 p_n-b_n\log_2 b_n\big].\end{aligned}
\label{eq:MI}
\end{equation}
The purity of $ABC$ gives $S_C=S_{AB}$ and $S_{AC}=S_B$. It follows that $I(A:B)+I(A:C)=2S_A=2$ bits for every $q$. In the Bell limit $I(A:B)=2$ and $I(A:C)=0$. As $q\to1$, each pair, $AB$ and $AC$, carries one bit of mutual information, while the Alice--Bob logarithmic negativity vanishes. Alice remains maximally entangled with the combined system $BC$ throughout.

The infinite sums admit explicit numerical error estimates. After summing Eq.~\eqref{eq:negativity} through $n=N$, the remaining contribution obeys
\begin{equation} 0\leq\mathcal N-\mathcal N^{(N)}\leq\frac{d\,q^{N+2}}{2[q^2+(N+1)d]}. \label{eq:Ntail}\end{equation}
For probabilities summed over $n=0,\ldots,K-1$, the exact tail masses are $t_p=q^K(1+Kd/2)$ and $t_b=q^{K-1}(q+Kd/2)$. The corresponding conditional means of the shifted tail index are
\begin{equation}
\begin{aligned} \mu_p&=\frac{q(3+Kd)}{d(2+Kd)},\qquad \mu_b=\frac{q(1+2q+Kd)}{d(2q+Kd)},\\ U(t,\mu)&=-t\log_2t+t\,g_2(\mu),\qquad g_2(\mu)=(1+\mu)\log_2(1+\mu)-\mu\log_2\mu.
\end{aligned}
\label{eq:entropytail}\end{equation}
A geometric distribution maximizes the entropy for a fixed mean occupation. Applying this fact to the normalized tail bounds its omitted entropy by $U(t,\mu)$. If $I^{(K)}$ denotes Eq.~\eqref{eq:MI} evaluated with the finite sums, the certified interval is $I^{(K)}-U(t_p,\mu_p)\leq I\leq I^{(K)}+U(t_b,\mu_b)$. The code selects the cutoff using both Eq.~\eqref{eq:Ntail} and these entropy bounds. For example, $q=0.5$ gives $E_{\mathcal N}=0.441378562536$ and $I=1.28853127578$, while $q=0.9$ gives $E_{\mathcal N}=0.0864730065373$ and $I=1.04528757551$.
\subsection{Amplifier channel and entanglement fidelity}

Tracing the interior in Eq.~\eqref{eq:isometry} defines a quantum limited amplifier \cite{Holevo2007,Caruso2006}. On the full bosonic oscillator, its Kraus representation is
\begin{equation}
\mathcal A_q(\rho)=\sum_{k=0}^\infty A_k\rho A_k^\dagger,\qquad A_k=\frac{q^{k/2}}{\sqrt{k!}}(b^\dagger)^k d^{(\hat N+1)/2},\qquad A_k\ket m=\sqrt{\binom{m+k}{k}}\,d^{(m+1)/2}q^{k/2}\ket{m+k}.
\label{eq:kraus}
\end{equation}
The identity $\sum_k\binom{m+k}{k}q^k=d^{-(m+1)}$ verifies $\sum_kA_k^\dagger A_k=\mathbb I$ for Eq.~\eqref{eq:kraus}. Its gain is $G_q=d^{-1}$. Every $A_k$ with $k>0$ raises Bob's occupation, giving an output orthogonal to the initial Bell vector. Only $A_0$ contributes to the Bell overlap, which gives
\begin{equation}
F_e=\bra{\Psi_{AB}}\rho_{AB}\ket{\Psi_{AB}}=\frac{(\sqrt d+d)^2}{4},\qquad \ket{\Psi_{AB}}=\frac{\ket{0,0}+\ket{1,1}}{\sqrt2}.
\label{eq:fidelity}
\end{equation}
The channel formulation connects the density matrix calculation to the fidelity analysis of Ref.~\cite{Sen2024Entanglement}. Writing $\mathcal G_G=\mathcal A_{1-1/G}$ for the channel labeled by its gain, amplifiers with $G_2\geq G_1\geq1$ obey $\mathcal G_{G_2}=\mathcal G_{G_2/G_1}\circ\mathcal G_{G_1}$. Increasing $q$ applies an additional local channel to the exterior subsystem. The monotonic decrease of accessible entanglement and mutual information follows from local channel monotonicity and the data processing inequality. All the scalar measures depend on the geometry through $q=\exp(-X)$ alone. For a fixed Killing frequency, $X=\omega/T_H=2\pi(M\omega)(1+s_h)^2/s_h$. An independent measurement of the mass at a known Killing frequency can thus relate the observed correlations to the smearing length.

\subsection{Frequency and position in the geometric comparison}
\label{sec:frequency}

The physical comparison is specified by the frequency held fixed. Combining Eqs.~\eqref{eq:tolman} and \eqref{eq:bogoliubov},
\begin{equation} q=\exp\!\left[-\frac{2\pi\omega}{\kappa}\right]=\exp\!\left[-\frac{2\pi\omega_{\rm loc}\sqrt{f(r_o)}}{\kappa}\right]=\exp\!\left[-\frac{\omega_{\rm loc}}{T_{\rm loc}}\right]. 
\label{eq:frequencyprotocols}
\end{equation}
A fixed Killing frequency selects the same outgoing mode as the observer position varies. A fixed local frequency instead selects $\omega=\omega_{\rm loc}\sqrt{f(r_o)}$ at each position. At fixed $M$ and $\omega$, Eqs.~\eqref{eq:kappamonotone} and \eqref{eq:negativity} show that increasing $\ell$ reduces $q$ and increases the accessible entanglement. For a fixed local frequency at an areal radius exterior to all compared horizons, the increasing lapse and decreasing surface gravity both reduce $q$.

A comparison at fixed proper distance organizes the local result differently. Equation~\eqref{eq:rindlerexpansion} gives
\begin{equation} 
q_{\varrho}=\exp\!\left\{-2\pi\omega_{\rm loc}\varrho\left[1+\frac{f''_+\varrho^2}{12}+O(\varrho^4)\right]\right\}. 
\label{eq:fixedproper}
\end{equation}
The leading factor depends only on $\omega_{\rm loc}\varrho$. The first curvature correction contains $f''_+=-1/(2M^2)+3\ell^2/(8M^4)+O(\ell^3/M^5)$, so its smearing dependence begins at order $\Theta$~\cite{MartinMartinez2010,Bruschi2010,Luongo2026Entanglement}.

The cancellation at fixed proper distance extends beyond the Rindler region. Set $r_o=M(1+s_h\cosh\xi)$ with $\xi>0$, and define $L_{\varrho}=\sqrt{f(r_o)}/\kappa$, so $q_{\varrho}=\exp(-2\pi\omega_{\rm loc}L_{\varrho})$. Equation~\eqref{eq:properdistance} gives
\begin{equation}
\frac{\varrho}{M}=\xi+s_h\sinh\xi,\qquad \frac{L_{\varrho}}{M}=\frac{(1+s_h)^2\sinh\xi}{1+s_h\cosh\xi}.\label{eq:fixedproperexact}
\end{equation}
At fixed $M$ and $\varrho$, the first identity gives $\partial_{s_h}\xi=-\sinh\xi/(1+s_h\cosh\xi)$. Substitution into the second yields
\begin{equation}
\left.\frac{1}{L_{\varrho}}\frac{\partial L_{\varrho}}{\partial s_h}\right|_{M,\varrho}=-\frac{(1-s_h)(\cosh\xi-1)[2+s_h(1+\cosh\xi)]}{(1+s_h)(1+s_h\cosh\xi)^2}.\label{eq:fixedpropersign}
\end{equation}
Since $s_h$ decreases as $\ell$ increases, $L_{\varrho}$ grows and $q_{\varrho}$ falls at any finite proper distance for fixed $\omega_{\rm loc}$. The derivative vanishes at the Schwarzschild value $s_h=1$. The first change in the ideal channel is therefore quadratic in $\ell$ even when the observer lies outside the range of the local expansion.

For fixed $M\omega$, Eq.~\eqref{eq:smalltheta} yields
\begin{equation}
q = q_{\rm S}\left[1-\frac{32\omega\Theta}{M}+O\!\left(\frac{(M\omega)\Theta^{3/2}}{M^3}\right)\right],\qquad q_{\rm S}=\e^{-8\pi M\omega}.
\label{eq:qsmalltheta}
\end{equation}
For each smooth channel observable $\mathcal Q(q)$, Eq.~\eqref{eq:qsmalltheta} gives $\mathcal Q-\mathcal Q_{\rm S}=-(32\omega\Theta/M)q_{\rm S}\mathcal Q^{\prime}(q_{\rm S})+O((\sqrt{\Theta}/M)^3)$ at fixed $M\omega$. For example, writing $d_{\rm S}=1-q_{\rm S}$ gives $F_e-F_{e,{\rm S}}=(8\omega\Theta/M)q_{\rm S}(1+\sqrt{d_{\rm S}})(1+2\sqrt{d_{\rm S}})+O((\sqrt{\Theta}/M)^3)$. The positive coefficient gives the leading gain in Bell fidelity.

One may also place the observer at a fixed fractional horizon offset, $r_o=(1+\epsilon_o)r_+$, with $\epsilon_o>0$. At fixed mass and local frequency, Eqs.~\eqref{eq:tolman} and \eqref{eq:kappa} give the exact thermal parameter
\begin{equation} 
\begin{aligned} q_{\epsilon_o}&=\exp\!\left[-2\pi\omega_{\rm loc}\sqrt{\mathcal F_{\epsilon_o}(s_h)}\right],\\ \mathcal F_{\epsilon_o}(s_h)&\equiv\frac{f((1+\epsilon_o)r_+)}{\kappa^2}=\frac{\epsilon_o M^2(1+s_h)^3[(\epsilon_o+2)s_h+\epsilon_o]}{(1+\epsilon_o)^2s_h^2}. 
\end{aligned} 
\label{eq:fractionaloffset}
\end{equation}
The stationary condition $(\epsilon_o+2)s_h^2-s_h-\epsilon_o=0$ gives $s_*=[1+\sqrt{1+4\epsilon_o(\epsilon_o+2)}]/[2(\epsilon_o+2)]$. The minimum of $\mathcal F_{\epsilon_o}$ maximizes $q_{\epsilon_o}$ and minimizes the accessible entanglement at $\eta_*=(1-s_*^2)/2$. For small offsets, $\eta_*=3/8-3\epsilon_o/8+O(\epsilon_o^2)$. The leading Rindler expression follows from $\mathcal F_{\epsilon_o}\simeq2\epsilon_o r_+/\kappa$ when $\epsilon_o r_+\ll\Delta_h$. The minimum arises because both the horizon radius and the temperature change along this sequence. Luongo \textit{et al.} found a related minimum using the static observer's proper acceleration in a Reissner--Nordstr\"om geometry~\cite{Luongo2026Entanglement}. Here the thermal parameter uses the Tolman scale $a_R=\kappa/\sqrt{f(r_o)}$. Equation~\eqref{eq:tolman} gives their ratio at finite radius and their common leading value close to the horizon.

Table~\ref{tab:fourprotocols} gives the first smearing correction for each prescription. Write $X=-\ln q$, let $X_{\rm S}$ be its value in the Schwarzschild reference geometry, and set $u_{\rm S}=M/R$, where $R>2M$ is the reference areal radius. At fixed $R$, the lapse is exactly $f(R)=1-2u_{\rm S}+2\eta u_{\rm S}^2$. At fixed proper distance, expanding Eq.~\eqref{eq:fixedpropersign} about $s_h=1$ gives the coefficient $1/4-u_{\rm S}^2$. It tends to zero as $R\to2M$ and approaches the fixed Killing frequency coefficient at large $R$. The fixed radius expansion assumes $\eta u_{\rm S}^2/(1-2u_{\rm S})\ll1$.
\begin{table}[htbp]\centering\small\setlength{\tabcolsep}{6pt}\begin{tabular}{lccccc}\toprule Quantities held fixed & Leading $(X-X_{\rm S})/X_{\rm S}$ & $q$ & $E_{\mathcal N}$ & $I(A:B)$ & $F_e$ \\ \midrule $\omega$ & $\eta^2/4$ & 0.229628 & 0.709054 & 1.554130 & 0.679043 \\ $\omega_{\rm loc},\ r_o$ & $\eta u_{\rm S}^2/(1-2u_{\rm S})$ & 0.001218 & 0.998244 & 1.993075 & 0.998173 \\ $\omega_{\rm loc},\ \varrho$ & $\eta^2(1/4-u_{\rm S}^2)$ & 0.283322 & 0.651482 & 1.490410 & 0.610935 \\ $\omega_{\rm loc},\ r_o/r_+$ & $-\eta/[4(1+\epsilon_o)]$ & 0.313425 & 0.620461 & 1.457607 & 0.573937 \\ \bottomrule\end{tabular}\caption{Frequency and position held fixed at a common mass. The analytic column gives the first term about $\eta=\ell/M=0$. The numerical columns use the exact effective lapse at $\eta=0.4$. All rows start from $M\omega=0.05$, $R/M=2.02$, and thus $\epsilon_o=0.01$ for the last row. The common reference has $E_{\mathcal N}=0.650139$ bits, $I(A:B)=1.488968$ bits, and $F_e=0.609335$.}\label{tab:fourprotocols}\end{table}

As $\Delta_h\to0^+$ at fixed $M$ and positive Killing frequency, the simple horizon family gives $q\to0$. At fixed $\omega_{\rm loc}$ and finite $\varrho$, Eq.~\eqref{eq:fixedproperexact} instead gives $\xi\to\varrho/M$ and $q_{\varrho}\to\exp[-2\pi M\omega_{\rm loc}\sinh(\varrho/M)]$. For $\varrho\ll M$, the exponent reduces to $-2\pi\omega_{\rm loc}\varrho$ at leading order. These limits describe sequences of observers and packets in nonextremal geometries. The mode construction at an exactly extremal horizon uses a different radial scaling~\cite{Luongo2026Entanglement}.

\section{A freely falling atom and acceleration radiation}\label{sec:falling_atom}
\subsection{Atomic motion and the outgoing optical phase}

\begin{figure}[!htbp]\centering\includegraphics[width=0.7\textwidth]{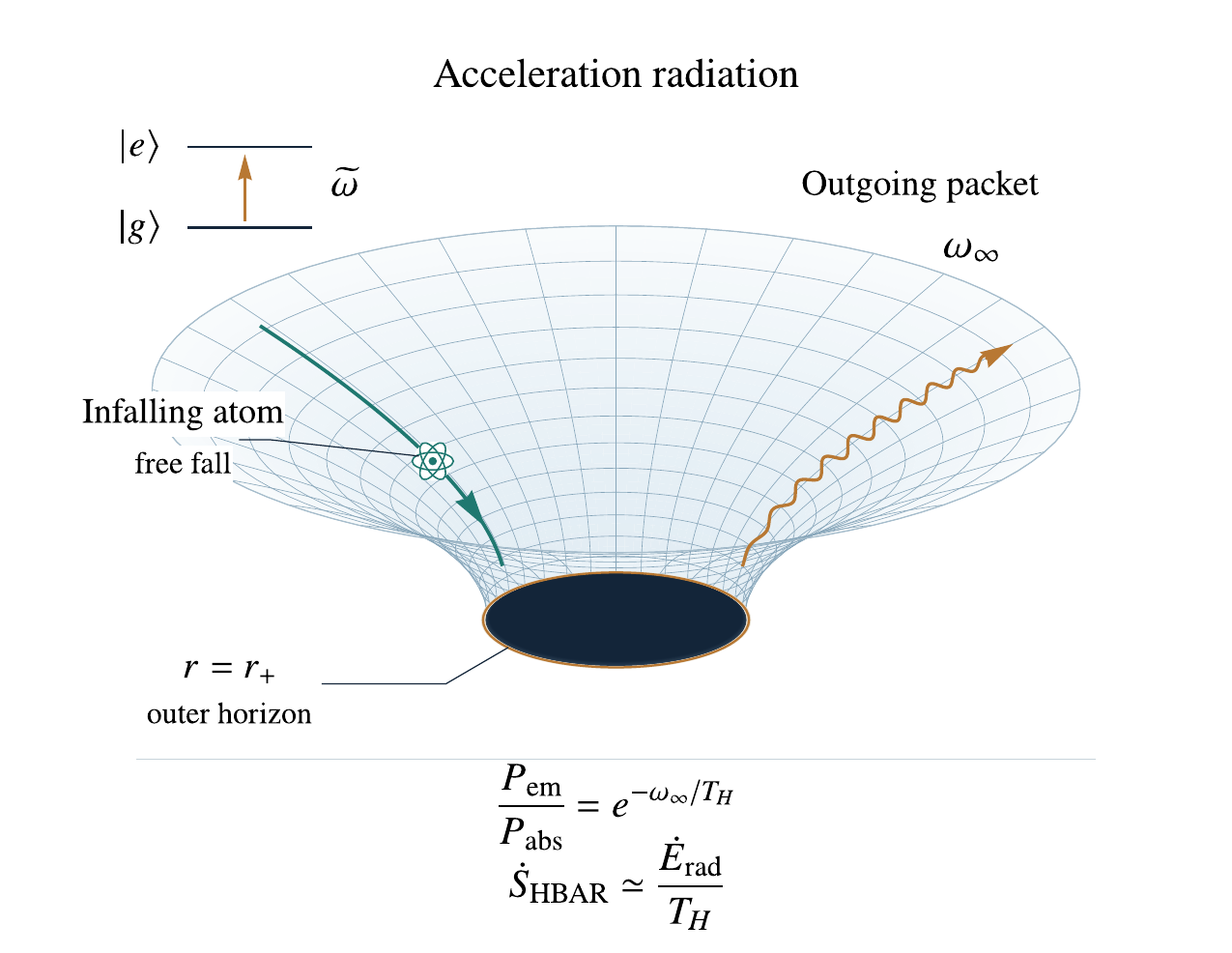}\caption{Freely falling two level atom coupled to an outgoing scalar packet in the prepared exterior Boulware vacuum. The labels $\widetilde\omega=\Omega$ and $\omega_\infty=\omega$ denote the proper atomic gap and Killing frequency. The upward atomic transition accompanies the creation of one outgoing quantum through the counterrotating interaction. The displayed detailed balance and entropy coefficient apply in the leading horizon and near thermal regimes, respectively.}\label{fig:lorentzian_hbar_setup}\end{figure}

We consider a two level atom in free fall with proper transition frequency $\Omega>0$, weakly coupled to an outgoing scalar mode of Killing frequency $\omega>0$. The entanglement calculation uses a state regular across the future horizon. For the atomic experiment, we prepare the selected exterior mode in its Boulware vacuum, following Refs.~\cite{Boulware1975,Scully2018,AziziMaster2021}. The outgoing phase varies along the atomic geodesic and enters its transition amplitude~\cite{UnruhWald1984,Camblong2020}. Figure~\ref{fig:lorentzian_hbar_setup} illustrates the excitation of the atom accompanied by the creation of an outgoing scalar quantum.

Let $\mathcal E$ denote the atom's conserved specific energy. Using $\mathcal E=f\dot t$ and the normalization of the four velocity, we obtain \begin{equation} \dot t=\frac{\mathcal E}{f},\qquad \dot r=-\sqrt{\mathcal E^2-f},\qquad \frac{\dd\tau}{\dd r}=-\frac{1}{\sqrt{\mathcal E^2-f}},\qquad \frac{\dd t}{\dd r}=-\frac{\mathcal E}{f\sqrt{\mathcal E^2-f}}. \label{eq:hbar_geodesics}\end{equation} Here a dot denotes differentiation with respect to proper time $\tau$, and the inward sign selects infall. For release from rest at infinity, $\mathcal E=1$ and $1-f=2M(r-\ell)/r^2$. With $x=r-r_+$, $y=\sqrt{r-\ell}$, $a_+=\sqrt{r_+-\ell}$, and the choice $\tau(r_+)=0$, integration gives 
\begin{equation} 
\tau(r)=-\sqrt{\frac{2}{M}}\frac{x}{y+a_+}\left[\frac{y^2+ya_++a_+^2}{3}+\ell\right]. 
\label{eq:hbar_tau_exact}
\end{equation} 
Equation~\eqref{eq:hbar_tau_exact} follows by integrating $r/\sqrt{2M(r-\ell)}$ and factoring the difference of two cubic powers. Its factored form also preserves numerical accuracy as $x$ approaches zero.
The outgoing phase is governed by the retarded coordinate $u=t-r_*$. Using $f=x(x+\Delta_h)/(r_++x)^2$ in Eq.~\eqref{eq:hbar_geodesics} gives 
\begin{equation} 
\begin{aligned} 
\tau(x)&=\tau_H-\frac{x}{\mathcal E}-\frac{\kappa x^2}{2\mathcal E^3}+O(x^3),\\ u(x)&=-\frac{1}{\kappa}\ln\frac{x}{L_0}+u_0+\mathfrak b\,x+O(x^2),\\ \mathfrak b&=\frac{f''(r_+)}{4\kappa^2}-\frac{1}{2\mathcal E^2}=\frac{2r_+(2r_--r_+)}{\Delta_h^2}-\frac{1}{2\mathcal E^2}. 
\end{aligned} 
\label{eq:hbar_phase_expansion}
\end{equation}
The positive reference length $L_0$ fixes an irrelevant phase origin. To obtain the regular coefficient in Eq.~\eqref{eq:hbar_phase_expansion}, expand $\dd u/\dd r=-2/f-1/(2\mathcal E^2)+O(f)$ and use $f=2\kappa x+f''(r_+)x^2/2+O(x^3)$. The regular coefficient contains both a geometric contribution and a term fixed by the atomic trajectory.

Along infall, $t$ contains $-(2\kappa)^{-1}\ln(x/L_0)$, while the outgoing radial phase contains $r_*=(2\kappa)^{-1}\ln(x/L_0)+O(1)$. Their difference doubles the coefficient in $u$. The ingoing coordinate $t+r_*$ is regular along the same trajectory. This distinction selects the outgoing sector responsible for the thermal term in the present experiment and connects its optical phase to the conformal mode analysis of Ref.~\cite{Camblong2020}. The radial support of the interaction lies where $\mathcal E^2>f$, so the geodesic measure remains real throughout the selected window.

An exact regular representation is useful for comparing this expansion with the switched detector response. Define $v(x)=\sqrt{\mathcal E^2-f(r_++x)}$ and write $u=-\kappa^{-1}\ln(x/L_0)+u_0+U(x)$. Rationalizing the square root in Eq.~\eqref{eq:hbar_geodesics} yields \begin{equation} 
\frac{\dd\tau}{\dd x}=-\frac{1}{v(x)},\qquad \frac{\dd U}{\dd x}=-2+\frac{2r_-^2}{\Delta_h(x+\Delta_h)}-\frac{1}{v(x)[\mathcal E+v(x)]},\qquad \tau(0)=U(0)=0. 
\label{eq:hbar_regular_phase}
\end{equation} 
Both derivatives in Eq.~\eqref{eq:hbar_regular_phase} are finite at the outer horizon, with $U'(0)=\mathfrak b$. These equations carry the full radial dependence of the selected effective metric into the detector integral and provide a stable alternative to evaluating coordinate times separately.
\subsection{Transition amplitudes and the thermal factor}

In the central frequency approximation, the outgoing mode is $\varphi_\omega=w(r)\e^{-i\omega u}$, where the real envelope satisfies $w(r_+)=1$. The coupling $g$ includes the mode normalization and its angular value along the trajectory. We use $w=r_+/r$ to include spherical dilution or $w=1$ for the leading local mode. For a narrow packet this prescription assumes that its envelope in retarded time varies slowly over the part of the trajectory contributing to the amplitude. A resolved packet envelope can be included in $\varphi_\omega$ before integration. With real switching function $\chi(\tau)$, the interaction Hamiltonian is
\begin{equation} 
H_I(\tau) = g\chi(\tau)\left[b\varphi_\omega(\tau)+b^\dagger\varphi_\omega^*(\tau)\right]\left[\sigma_-\e^{-i\Omega\tau}+\sigma_+\e^{i\Omega\tau}\right].
\label{eq:hbar_interaction}
\end{equation} 
The counterrotating term $b^\dagger\sigma_+$ in Eq.~\eqref{eq:hbar_interaction} excites the atom while creating a field quantum. The term $b\sigma_+$ excites the atom while absorbing one quantum. These processes give the single quantum probabilities that enter the field master equation \cite{Scully2018,AziziMaster2021}.
To order $g^2$, the emission and absorption probabilities are 
\begin{equation}
\begin{aligned} 
P_{\rm em}&=g^2\left|\int_{r_+}^{\infty}\frac{\chi(r)w(r)}{\sqrt{\mathcal E^2-f(r)}}\e^{i\Omega\tau(r)+i\omega u(r)}\dd r\right|^2,\\ P_{\rm abs}&=g^2\left|\int_{r_+}^{\infty}\frac{\chi(r)w(r)}{\sqrt{\mathcal E^2-f(r)}}\e^{i\Omega\tau(r)-i\omega u(r)}\dd r\right|^2. 
\end{aligned} 
\label{eq:hbar_exact_probabilities}
\end{equation} 
Here $\chi(r)=\chi[\tau(r)]$ restricts the interaction to the exterior trajectory. Occupation factors $n+1$ and $n$ multiply these probabilities for an initial field state with $n$ quanta. Thus absorption probes an initially occupied mode, while emission is possible from the vacuum.

The radial envelope specifies the local approximation to the field mode. For a massless scalar with multipole number $j$, the radial potential is $V_j=f[j(j+1)/r^2+f'/r]$. It approaches zero at a simple horizon, giving outgoing phases proportional to $\e^{-i\omega u}$. A window satisfying $|V_j|\ll\omega^2$ isolates this sector. More explicitly, $V_j(r_++x)=2\kappa U_{j,+}x+O(x^2)$, where $U_{j,+}=j(j+1)/r_+^2+2\kappa/r_+$. The endpoint scale $x_{\rm dom}\sim\mathcal E\max(1,s)/\Omega$, with $s=\omega/\kappa$, therefore has the optical control parameter $2\kappa U_{j,+}x_{\rm dom}/\omega^2$. Propagation through the exterior potential introduces additional scattering amplitudes for a mode normalized at infinity~\cite{Anacleto2020Scattering}.

For the horizon limit, we take a smooth window with $\chi(\tau_H)=1$ that varies slowly on the scale $x_{\rm dom}$ and decays smoothly at its outer end. For any other nonzero value at the horizon, both leading probabilities acquire the common factor $|\chi(\tau_H)|^2$. Insert Eq.~\eqref{eq:hbar_phase_expansion} and define $p=\Omega/\mathcal E$. The leading emission amplitude is proportional to $\mathcal E^{-1}\int_0^\infty(x/L_0)^{-is}\e^{-ipx}\dd x$. With a positive regulator $\epsilon$, we obtain 
\begin{equation} 
\begin{aligned} 
J_s(\epsilon,p)&=\int_0^\infty\left(\frac{x}{L_0}\right)^{-is}\e^{-(\epsilon+ip)x}\dd x=\frac{L_0^{is}\Gamma(1-is)}{(\epsilon+ip)^{1-is}},\\ |\Gamma(1-is)|^2&=\frac{\pi s}{\sinh(\pi s)},\qquad \lim_{\epsilon\to0^+}|J_s(\epsilon,p)|^2=\frac{2\pi s}{p^2(\e^{2\pi s}-1)}. 
\end{aligned} 
\label{eq:hbar_mellin}
\end{equation}
The phase of $(\epsilon+ip)^{1-is}$ supplies the exponential factor in Eq.~\eqref{eq:hbar_mellin}. Absorption follows by changing $s$ to $-s$. Moreover, the factor $p^{-2}=\mathcal E^2/\Omega^2$ cancels the geodesic measure $\mathcal E^{-2}$. Writing $P_{\rm em}^{(0)}$ and $P_{\rm abs}^{(0)}$ for the resulting probabilities, we find
\begin{equation} 
P_{\rm em}^{(0)}=\frac{2\pi g^2\omega}{\kappa\Omega^2}\frac{1}{\e^{2\pi\omega/\kappa}-1},\qquad P_{\rm abs}^{(0)}=\frac{2\pi g^2\omega}{\kappa\Omega^2}\frac{1}{1-\e^{-2\pi\omega/\kappa}},\qquad \frac{P_{\rm em}^{(0)}}{P_{\rm abs}^{(0)}}=\e^{-\omega/T_H}.
\label{eq:hbar_single_probabilities}
\end{equation}
Equation~\eqref{eq:hbar_single_probabilities} displays the same near horizon detailed balance found in the conformal analysis of generalized static geometries \cite{Camblong2020,AziziMaster2021}. At fixed $M$, $\omega$, $\Omega$, and common normalization, a smaller surface gravity suppresses $P_{\rm em}^{(0)}$. To display the change in both probabilities, let $X_{\rm S}=8\pi M\omega$, $n_{\rm S}=(\e^{X_{\rm S}}-1)^{-1}$, and $\delta=\ell^2/(4M^2)=4\Theta/(\pi M^2)$. Then $X=\omega/T_H=X_{\rm S}(1+\delta)+O(\eta^3)$. Using $\dd\bar n_\beta/\dd X=-\bar n_\beta(1+\bar n_\beta)$ in Eq.~\eqref{eq:hbar_single_probabilities} gives
\begin{equation}\begin{aligned} \frac{\Omega^2 P_{\rm em}^{(0)}}{g^2}&=X_{\rm S}n_{\rm S}\big\{1+[1-X_{\rm S}(1+n_{\rm S})]\delta\big\}+O(\eta^3),\\ \frac{\Omega^2 P_{\rm abs}^{(0)}}{g^2}&=X_{\rm S}(1+n_{\rm S})\big\{1+[1-X_{\rm S}n_{\rm S}]\delta\big\}+O(\eta^3).\end{aligned}\label{eq:atomic_smalltheta}\end{equation}
For $X_{\rm S}>0$, $X_{\rm S}(1+n_{\rm S})>1$ and $X_{\rm S}n_{\rm S}<1$. Thus smearing suppresses spontaneous emission and enhances absorption from an occupied mode at this order. The initial specific energy cancels from both coefficients. Studies of quantum corrected, charged, and generalized regular metrics provide complementary applications of this mechanism \cite{Sen2022HBAR,Jana2024HBAR,Jana2025HBAR,Uktamov2026HBAR}.

Setting $\ell=0$ gives $\kappa=1/(4M)$ and $P_{\rm em}^{(0)}=8\pi Mg^2\omega/[\Omega^2(\e^{8\pi M\omega}-1)]$, the radial horizon expression underlying the original HBAR calculation \cite{Scully2018}. The common normalization cancels from the emission to absorption ratio. Below, $P_{\rm em}$ and $P_{\rm abs}$ in the leading rate equations denote the probabilities in Eq.~\eqref{eq:hbar_single_probabilities}.
\subsection{Finite switching and the approach to extremality}

Restricting the regulated Mellin integral to $x_1<x<x_2$ gives $J_s(x_1,x_2;\epsilon,p)=L_0^{is}(\epsilon+ip)^{-1+is}[\gamma(1-is,(\epsilon+ip)x_2)-\gamma(1-is,(\epsilon+ip)x_1)]$, where $\gamma$ is the lower incomplete gamma function. An abrupt outer cutoff contributes a boundary amplitude of order $p^{-1}$, the same order as the horizon amplitude. The smooth outer switching assumed above suppresses this contribution. For $\chi=\e^{-x/L}$, set $h=(pL)^{-1}$. Keeping the leading optical phase, proper time, and radial amplitude in Eq.~\eqref{eq:hbar_mellin} gives $P_{\rm em}^{(L)}/P_{\rm em}^{(0)}=\e^{2s\arctan h}/(1+h^2)$ and $(P_{\rm em}^{(L)}/P_{\rm abs}^{(L)})/q=\e^{4s\arctan h}$, where $q=\e^{-2\pi s}$. These expressions isolate the effect of the switching.

The first correction for the full trajectory has a simple form in proper time. Let $\lambda=\mathcal E(\tau_H-\tau)$ and $W(x)=\chi[\tau(r_++x)]w(r_++x)$, with $W(0)=1$. Equation~\eqref{eq:hbar_phase_expansion} gives $x=\lambda-\kappa\lambda^2/(2\mathcal E^2)+O(\lambda^3)$. On substituting in $u$, the logarithm contributes $\lambda/(2\mathcal E^2)$, canceling the trajectory term in $\mathfrak b\,\lambda$. We obtain $u=-\kappa^{-1}\ln(\lambda/L_0)+u_0+f''(r_+)\lambda/(4\kappa^2)+O(\lambda^2)$, while the integration measure is $\dd\lambda/\mathcal E$. The linear term in the emission integral uses the first Mellin moment, whose ratio to the zeroth moment is $(1-is)/(ip)$. With $W'(0)=\dd W/\dd x|_{x=0}$, this gives
\begin{equation}
\frac{P_{\rm em}^{\rm sw}}{P_{\rm em}^{(0)}}=1+\frac{2s}{p}\left[\frac{f''(r_+)}{4\kappa}-W'(0)\right]+O\!\left((pM)^{-2}\right).
\label{eq:hbar_switched_correction}
\end{equation}
Let $D=(2s/p)[f^{\prime\prime}(r_+)/(4\kappa)-W^{\prime}(0)]$. The corresponding absorption probability and rate ratio are
\begin{equation}
\frac{P_{\rm abs}^{\rm sw}}{P_{\rm abs}^{(0)}}=1-D+O((pM)^{-2}),\qquad \frac{P_{\rm em}^{\rm sw}}{P_{\rm abs}^{\rm sw}}=\e^{-2\pi s}\big[1+2D+O((pM)^{-2})\big].\label{eq:hbar_switched_ratio}
\end{equation} This expansion holds at fixed $s$, geometry, trajectory, and window shape and width as $pM$ increases. For $\chi=\e^{-x/L}$ and $w=r_+/r$, the bracket in Eq.~\eqref{eq:hbar_switched_correction} reduces to $1/L+r_-/(r_+\Delta_h)$. The first term measures switching and the second measures the combined effect of spherical dilution and the regular optical phase. For small smearing, $r_-/(r_+\Delta_h)=\ell/(4M^2)+O(\ell^2/M^3)$. At fixed $M,\omega,\Omega,\mathcal E,L$, and horizon normalization, this profile contributes $s_{\rm S}\ell/(2pM^2)$ to the relative emission probability at first order in $p^{-1}$ and $\ell$, where $s_{\rm S}=4M\omega$. The linear contribution comes from the slope of $w=r_+/r$ as the horizon moves. The leading thermal probability first changes at order $\ell^2\propto\Theta$.

For a smooth interaction of radial width $L$, the scale requirements are $L\ll\Delta_h,r_+$ and $pL\gg\max(1,s)$. In the radial phase the regular term shifts the linear frequencies to $p_{\rm em}=p-\omega\mathfrak b$ and $p_{\rm abs}=p+\omega\mathfrak b$. Their common leading value applies for $p\gg\omega|\mathfrak b|$. As $\Delta_h$ decreases, the first two conditions imply $p\Delta_h\gg\max(1,s)$. At fixed $\omega>0$, $s=\omega/\kappa$ grows near extremality and the required gap increases. The double root has a distinct optical expansion~\cite{Luongo2026Entanglement}. At large $s$, the leading emission probability is exponentially small, which makes its relative change particularly sensitive to the switching.

For numerical evaluation, set $x=L_0\e^z$ in Eq.~\eqref{eq:hbar_exact_probabilities} and use Eq.~\eqref{eq:hbar_regular_phase}. The emission integrand becomes $x\chi(x)w(x)\e^{i\Omega\tau(x)-isz+i\omega U(x)}/v(x)$, which is smooth in $z$ and decays as $z\to-\infty$. The absorption phase reverses both terms proportional to $\omega$. Truncating the lower endpoint at $x_{\min}>0$ omits an amplitude of magnitude at most $|g|x_{\min}\max_{[0,x_{\min}]}|\chi w|/\min_{[0,x_{\min}]}v$. It is important to mention that the same cutoff and window are used for both processes.


\section{Field dynamics and HBAR entropy}\label{sec:hbar_dynamics}
\subsection{Random atomic injection and the stationary distribution}

Consider a dilute beam of ground state atoms injected at random coordinate times with mean rate $\mathcal R_{\rm at}$. Successive atoms interact independently with the same exterior oscillator. Its frequency and coupling remain fixed, and leakage is negligible on the relaxation time. We assume that the arrival times sample the field phase uniformly, so that phase sensitive terms average to zero. The probabilities per encounter are $(n+1)P_{\rm em}$ and $nP_{\rm abs}$ for an oscillator initially in $\ket n$. Both must be small over the appreciably populated occupations. With $\Gamma_{\rm em}=\mathcal R_{\rm at}P_{\rm em}$ and $\Gamma_{\rm abs}=\mathcal R_{\rm at}P_{\rm abs}$, the weak coupling master equation is~\cite{Scully2018,AziziMaster2021,Lindblad1976}
\begin{equation} 
\dot\rho=\Gamma_{\rm em}\mathcal D[b^\dagger]\rho+\Gamma_{\rm abs}\mathcal D[b]\rho,\qquad \mathcal D[L]\rho=L\rho L^\dagger-\frac12\{L^\dagger L,\rho\}. 
\label{eq:hbar_master}
\end{equation} 

The time in Eq.~\eqref{eq:hbar_master} is the asymptotic coordinate time used to define the injection rate. Taking diagonal matrix elements, with $p_n=\bra n\rho\ket n$, gives
\begin{equation} 
\dot p_n=\Gamma_{\rm em}[n p_{n-1}-(n+1)p_n]+\Gamma_{\rm abs}[(n+1)p_{n+1}-np_n]. 
\label{eq:hbar_populations}
\end{equation} 

The probability current from occupation $n$ to $n+1$ is $J_{n+1}=(n+1)(\Gamma_{\rm em}p_n-\Gamma_{\rm abs}p_{n+1})$. Equation~\eqref{eq:hbar_populations} gives $\dot p_n=J_n-J_{n+1}$, with $J_0=0$. At stationarity all currents vanish, so $p_{n+1}/p_n=\Gamma_{\rm em}/\Gamma_{\rm abs}$. Using the leading horizon probabilities in Eq.~\eqref{eq:hbar_single_probabilities} and summing the geometric sequence gives
\begin{equation} 
\rho_\beta=(1-q_\omega)\sum_{n=0}^{\infty}q_\omega^n\ket n\bra n,\qquad q_\omega=\frac{\Gamma_{\rm em}}{\Gamma_{\rm abs}}=\e^{-\beta_H\omega},\qquad \bar n_\beta=\frac{q_\omega}{1-q_\omega}=\frac{1}{\e^{\beta_H\omega}-1}. 
\label{eq:hbar_stationary}
\end{equation}

Here $\beta_H=1/T_H$. The temperature of the stationary mode is fixed by the emission to absorption ratio of a single atom. For finite switching, a ratio $0<R_{\rm sw}=P_{\rm em}^{\rm sw}/P_{\rm abs}^{\rm sw}<1$ gives the same geometric distribution with $q_\omega$ replaced by $R_{\rm sw}$. Its inverse temperature $-\ln R_{\rm sw}/\omega$ can depend on the packet. In what follows we use the leading horizon rates.

\subsection{Exact transient occupation and entropy}

Multiplying Eq.~\eqref{eq:hbar_populations} by $n$ and summing gives $\dot{\bar n}=\Gamma_{\rm em}(\bar n+1)-\Gamma_{\rm abs}\bar n$. With $\gamma=\Gamma_{\rm abs}-\Gamma_{\rm em}>0$, the solution is
\begin{equation}
\bar n(t)=\bar n_\beta+[\bar n(0)-\bar n_\beta]\e^{-\gamma t},\qquad \bar n(0)=0\ \Longrightarrow\ \bar n(t)=\bar n_\beta(1-\e^{-\gamma t}). 
\label{eq:hbar_occupation}
\end{equation} 

The complete distribution for the exterior vacuum initial state follows from the generating function $G(z,t)=\sum_{n\geq0}p_n(t)z^n$, with $G(z,0)=1$. Multiplying Eq.~\eqref{eq:hbar_populations} by $z^n$ and summing gives the evolution equation and its vacuum solution,
\begin{equation}
\partial_tG=(z-1)[\Gamma_{\rm em}G+(\Gamma_{\rm em}z-\Gamma_{\rm abs})\partial_zG],\qquad G(z,t)=\frac{1}{1+\bar n(t)(1-z)}.
\label{eq:hbar_generating_function}
\end{equation} 

Substituting the second expression into the first recovers $\dot{\bar n}=\Gamma_{\rm em}-\gamma\bar n$. Expanding in $z$ gives $p_n(t)=[1+\bar n(t)]^{-1}[\bar n(t)/(1+\bar n(t))]^n$. Thus the initially empty oscillator remains geometric throughout the evolution. In the leading horizon approximation, Eq.~\eqref{eq:hbar_single_probabilities} gives $\gamma=2\pi\mathcal R_{\rm at}g^2\omega/(\kappa\Omega^2)$.

For this state, the von Neumann entropy and its derivative at $t>0$ are
\begin{equation} 
S_\omega(t)=[1+\bar n(t)]\ln[1+\bar n(t)]-\bar n(t)\ln\bar n(t),\qquad \dot S_\omega=\ln\frac{1+\bar n}{\bar n}\,\dot{\bar n}. 
\label{eq:hbar_transient_entropy}
\end{equation} 

The continuous vacuum limit gives $S_\omega(0)=0$. At early times, $\bar n(t)=\Gamma_{\rm em}t+O(t^2)$, so the logarithmic divergence of $\dot S_\omega$ at $t=0$ is integrable. For $E_\omega=\omega\bar n$, the instantaneous slope $\dd S_\omega/\dd E_\omega=\omega^{-1}\ln[(1+\bar n)/\bar n]$ approaches $\beta_H$ as the mode reaches equilibrium.

For the generator $\mathcal L$ in Eq.~\eqref{eq:hbar_master}, define $\dot Q_\omega=\Tr(H_\omega\mathcal L\rho)=\omega\dot{\bar n}$ with $H_\omega=\omega b^\dagger b$. Using natural logarithms for the field entropy, the relative entropy $D(\rho\Vert\rho_\beta)=\Tr[\rho(\ln\rho-\ln\rho_\beta)]$ gives the exact balance~\cite{Spohn1978}
\begin{equation} 
\dot S_\omega=\beta_H\dot Q_\omega+\sigma_\omega,\qquad \sigma_\omega=-\Tr\{\mathcal L\rho[\ln\rho-\ln\rho_\beta]\}\geq0,\qquad \frac{\dd}{\dd t}D(\rho\Vert\rho_\beta)=-\sigma_\omega.
\label{eq:hbar_entropy_balance}
\end{equation} 

For the geometric trajectory, $\sigma_\omega=\{\ln[(1+\bar n)/\bar n]-\beta_H\omega\}\dot{\bar n}$. For the vacuum solution, write $z_t=\exp(-\gamma t)$ and $X=\omega/T_H$. Since $\bar n=(1-z_t)/(\e^X-1)$, the exact entropy rate and its two contributions for $t>0$ are
\begin{equation}\begin{aligned} \dot S_\omega&=\frac{\gamma z_t}{\e^X-1}\ln\!\frac{\e^X-z_t}{1-z_t},\qquad \beta_H\dot E_\omega=\frac{\gamma X z_t}{\e^X-1},\\ \sigma_\omega&=\frac{\gamma z_t}{\e^X-1}\ln\!\frac{1-\e^{-X}z_t}{1-z_t}.\end{aligned}\label{eq:hbar_entropy_closed}\end{equation}
Since $0<\e^{-X}<1$ and $0<z_t<1$, the logarithm in $\sigma_\omega$ is positive. Integrating Eq.~\eqref{eq:hbar_entropy_balance} over the entire relaxation gives $\int_0^\infty\sigma_\omega\dd t=D(\ket0\bra0\Vert\rho_\beta)=-\ln(1-q_\omega)$. The final entropy therefore exceeds $\beta_H\omega\bar n_\beta$ by $\ln(1+\bar n_\beta)$, precisely the entropy produced during preparation.

\subsection{The near thermal entropy flux}

The order of the irreversible term determines when the HBAR entropy and energy increments are proportional. Set $\delta n=\bar n-\bar n_\beta$ at fixed $\omega$ and $T_H$. Since $\dot{\bar n}=-\gamma\delta n$ and $\ln[(1+\bar n_\beta)/\bar n_\beta]=\beta_H\omega$, expanding Eq.~\eqref{eq:hbar_entropy_balance} gives $\sigma_\omega=\gamma(\delta n)^2/[\bar n_\beta(1+\bar n_\beta)]+O((\delta n)^3)$. The entropy production starts at second order, while $\dot Q_\omega=-\omega\gamma\delta n$ is first order. For independent outgoing modes near their stationary occupations, summing the leading contributions yields~\cite{AziziThermodynamics2021}
\begin{equation}
\dot S_{\rm HBAR}\simeq\beta_H\sum_j\omega_j\dot{\bar n}_j\equiv\frac{\dot E_{\rm rad}}{T_H}, 
\label{eq:hbar_entropy_flux}
\end{equation}
where $E_{\rm rad}=\sum_j\omega_j\bar n_j$ is the energy in the selected exterior modes. At stationarity both sides of Eq.~\eqref{eq:hbar_entropy_flux} vanish. During relaxation the equation describes the change in their stored energy and entropy. An energy balance for the outgoing packets relates this increment to escaping radiation. A model of continuous extraction would give the corresponding luminosity.

The same incremental relation enters the bumblebee calculation of Ref.~\cite{AraujoLiu2025} and studies of generalized uncertainty and regular geometries~\cite{OvgunPantig2026HBAR,Uktamov2026HBAR}. For the present metric, the Lorentzian correction enters the stationary occupation through $\kappa$ and the finite interaction through $\tau$, $U$, and $w$.


\section{Horizon energy balance and the area correspondence}\label{sec:hbar_area}
\subsection{The mass dependent effective source}

For the Einstein gravitational action with smearing in the matter source, the Wald entropy is $S_H=A/4=\pi r_+^2$~\cite{Bekenstein1973,Wald1993,IyerWald1994}. To identify the energy variation conjugate to this entropy, first treat the coefficient $C$ in $f=1-2M/r+C/r^2$ as independent of $M$. The horizon equation gives $M=r_+/2+C/(2r_+)$. Since $T_H=(1-C/r_+^2)/(4\pi r_+)$, differentiation yields
\begin{equation}
\dd M=\frac12\left(1-\frac{C}{r_+^2}\right)\dd r_++\frac{\dd C}{2r_+}=T_H\dd\left(\frac A4\right)+\frac{\dd C}{2r_+}. \label{eq:hbar_horizon_variation}
\end{equation}
The mass function also gives a direct interpretation of the last term. The matter outside a sphere of radius $R$ contributes $E_{\rm out}(R)=M-m(R)=C/(2R)$ to the mass at infinity. Changing $C$ at fixed $R=r_+$ changes this contribution by $\dd C/(2r_+)$. When the horizon moves, the total variation is $\dd E_{\rm out}(r_+)=\dd C/(2r_+)-C\,\dd r_+/(2r_+^2)$. The second term accounts for the change in the lower limit of the exterior region.

In the effective Lorentzian family, $C=2M\ell$, so $\dd C=2\ell\dd M+2M\dd\ell$. Substitution into Eq.~\eqref{eq:hbar_horizon_variation} gives
\begin{equation}
T_H\dd\left(\frac A4\right)=\left(1-\frac{\ell}{r_+}\right)\dd M-\frac{M}{r_+}\dd\ell,\qquad T_H\dd\left(\frac A4\right)=\mathcal C_+\dd M\quad(\dd\ell=0),\qquad \mathcal C_+=1-\frac{\ell}{r_+}. \label{eq:hbar_source_factor}
\end{equation}
The same factor follows from the matter variation formula of Ma and Zhao~\cite{MaZhao2014}. With $T^0{}_0=-M\ell/(4\pi r^4)$, the integral at fixed $\ell$ gives $1+4\pi\int_{r_+}^{\infty}r^2\partial_M T^0{}_0\,\dd r=1-\ell/r_+$. For a simple outer horizon with $\ell>0$, one has $1/2<\mathcal C_+<1$. In the Schwarzschild limit, $\mathcal C_+=1$ and the mass variation equals the horizon heat.

The result extends to the complete smeared profile. For $m(r)=MF(r)$ with $F$ fixed, the horizon relation $M=r_+/[2F(r_+)]$ gives $\dd M/\dd r_+=[F-r_+F']/[2F^2]$. The surface gravity gives $T_H=[F-r_+F']/[4\pi r_+F]$, with $F$ and $F'$ evaluated at $r_+$. Combining the two expressions yields $T_H\dd(A/4)=F(r_+)\dd M$. At fixed smearing length, the fraction of the ADM mass variation that enters the horizon heat thus equals the enclosed mass fraction $F(r_+)$.
\subsection{Heat exchange and the chosen quasistatic path}

For outgoing modes close to their stationary occupations, Eq.~\eqref{eq:hbar_entropy_flux} gives the explicit spectral result $\dot S_{\rm HBAR}\simeq[2\pi M(1+s_h)^2/s_h]\sum_j\omega_j\dot{\bar n}_j$. To express it as an area rate, define $\dot A_p\equiv-\dot A_H^{\rm rad}>0$ for an outward radiative energy transfer $\dot E_{\rm rad}>0$, following the positive area loss convention of Ref.~\cite{AraujoLiu2025}. An atom can emit into the exterior field before crossing the horizon. Its remaining ingoing energy contributes separately to the total mass and area change~\cite{Scully2018,AziziThermodynamics2021}.

Let $\dot A_p^{(Q)}$ denote the area loss obtained by drawing the outward energy from horizon heat, $\dot Q_H^{\rm rad}=-\dot E_{\rm rad}$, with $\dd Q_H=T_H\dd(A/4)$. Let $\dot A_p^{(M)}$ denote the area loss obtained by assigning the same outward energy to ADM mass loss at fixed $\ell$, $\dot M^{\rm rad}=-\dot E_{\rm rad}$. The horizon equation gives $\mathcal C_+=1-\ell/r_+=r_+/(2M)=(1+s_h)/2$. Using this identity and $T_H=\Delta_h/(4\pi r_+^2)$ gives
\begin{equation}\begin{aligned}
\dot A_p^{(Q)}&=\frac{4}{T_H}\dot E_{\rm rad}=\frac{16\pi r_+^2}{\Delta_h}\dot E_{\rm rad},\\
\dot A_p^{(M)}&=\frac{4\mathcal C_+}{T_H}\dot E_{\rm rad}=\frac{16\pi r_+(r_+-\ell)}{\Delta_h}\dot E_{\rm rad}.
\end{aligned}\label{eq:positive_area_rates}\end{equation}
The second coefficient is also $(\partial A/\partial M)_\ell$, obtained directly by differentiating $A=4\pi r_+^2$. For the same emitted energy, assigning it to ADM mass loss gives a smaller radiative area loss by the factor $\mathcal C_+$. The entropy rates are
\begin{equation}\begin{aligned}
\dot S_{\rm HBAR}&\simeq\frac14\dot A_p^{(Q)}, &&\dot Q_H^{\rm rad}=-\dot E_{\rm rad},\\
\dot S_{\rm HBAR}&\simeq\frac{\dot A_p^{(M)}}{4\mathcal C_+}=\frac{\dot A_p^{(M)}}{2[1+\sqrt{1-2\ell/M}]}, &&\dot\ell=0,\quad\dot M^{\rm rad}=-\dot E_{\rm rad}.
\end{aligned}\label{eq:hbar_area_paths}\end{equation}
Both area rates reduce to $32\pi M\dot E_{\rm rad}$ in the Schwarzschild limit. At $\ell/M=0.4$, the fixed $\ell$ ADM assignment gives $\dot S_{\rm HBAR}/\dot A_p^{(M)}\simeq0.345492$, compared with $1/4$ for the heat assignment. In the bumblebee calculation, the corresponding coefficient depends on the chosen vacuum~\cite{AraujoLiu2025}. Here, it depends on the instantaneous horizon radius even at fixed $\ell$.

Along the fixed $\ell$ path, $M=r_+^2/[2(r_+-\ell)]$ and $T_H=(r_+-2\ell)/[4\pi r_+(r_+-\ell)]$. These relations give $(\dd M/\dd r_+)/T_H=2\pi r_+^2/(r_+-\ell)$. Integration yields
\begin{equation}
\mathcal S_M\equiv\int\frac{\dd M}{T_H}=\pi r_+^2+2\pi\ell r_++2\pi\ell^2\ln\frac{r_+-\ell}{L_*}+\mathrm{constant},\qquad \dd\ell=0. \label{eq:hbar_mass_integral}
\end{equation}
For a purely radiative, quasistatic decrease from $r_i$ to $r_f>2\ell$, the positive area loss is $A_p=4\pi(r_i^2-r_f^2)$. The heat assignment integrates to $\Delta S_{\rm HBAR}^{(Q)}\simeq A_p/4$. Along the fixed $\ell$ ADM path, $\Delta S_{\rm HBAR}^{(M)}\simeq\mathcal S_M(r_i)-\mathcal S_M(r_f)$, giving
\begin{equation}
\Delta S_{\rm HBAR}^{(M)}(A_p)\simeq\frac{A_p}{4}+2\pi\ell(r_i-r_f)+2\pi\ell^2\ln\!\frac{r_i-\ell}{r_f-\ell},\qquad r_f=\sqrt{r_i^2-\frac{A_p}{4\pi}}.\label{eq:hbar_area_integrated}
\end{equation}
Differentiation with respect to $A_p$ at fixed $r_i$ recovers $1/[4(1-\ell/r_f)]$. The integration follows a sequence of near thermal increments, with horizon radius and temperature evaluated at each step.
The arbitrary length $L_*>0$ cancels from the finite entropy increment. The linear and logarithmic terms in $\mathcal S_M$ arise from integrating the change in source normalization along the mass path. The gravitational entropy of the same configuration is $S_H=\pi r_+^2$. This energy assignment specifies the comparison with HBAR entropy corrections in other geometries~\cite{Sen2022HBAR,Jana2024HBAR,Jana2025HBAR}.

At fixed $M$, the area coefficient in the second line of Eq.~\eqref{eq:hbar_area_paths} contains $\mathcal C_+^{-1}=1+\ell/(2M)+O(\ell^2/M^2)$, while $T_H/T_{\rm Schw}=1-\ell^2/(4M^2)+O(\ell^3/M^3)$. The changing exterior source contributes at order $\sqrt\Theta$ to the area assigned to a given ADM energy loss. The thermal occupation at a fixed Killing frequency first changes at order $\Theta$.


\section{Parameter dependence and numerical results}\label{sec:results}
\subsection{Geometry and the frequency prescription}

\begin{figure}[htbp]\centering\includegraphics[width=\textwidth]{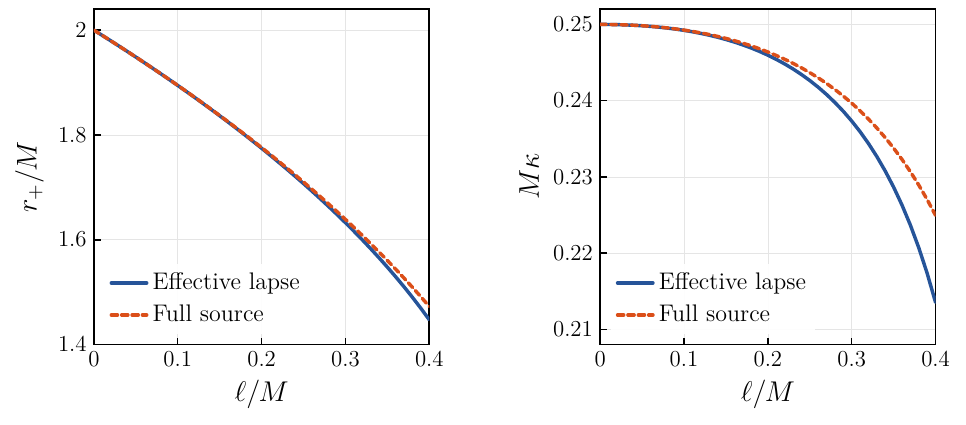}\caption{Outer horizon and surface gravity for the effective lapse and the complete Lorentzian source, at fixed mass. The effective quantities follow Eqs.~\eqref{eq:horizons} and \eqref{eq:kappa}. The complete source is evaluated from Eq.~\eqref{eq:lorentzianmass}. The horizontal variable is the independent ratio $\ell/M$.}\label{fig:geometry}\end{figure}

\begin{figure}[htbp]\centering\includegraphics[width=\textwidth]{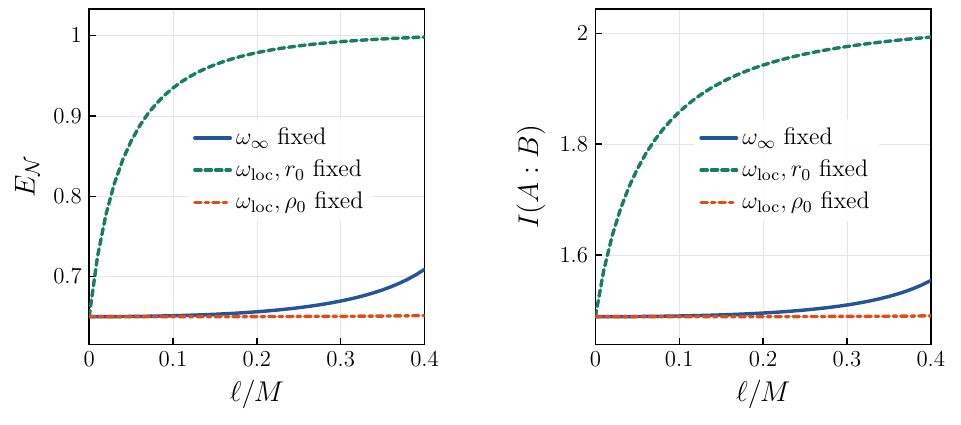}\caption{Logarithmic negativity and mutual information for three selections of the outgoing packet frequency. The common Schwarzschild reference has $M\omega=0.05$ and $r_o/M=2.02$. The local frequency is fixed by Eq.~\eqref{eq:tolman} at that reference point. The fixed distance curve uses the corresponding proper distance from Eq.~\eqref{eq:properdistance}. These curves compare ideal horizon resources with the stated Tolman frequency selections.}\label{fig:protocols}\end{figure}

\begin{figure}[htbp]\centering\includegraphics[width=\textwidth]{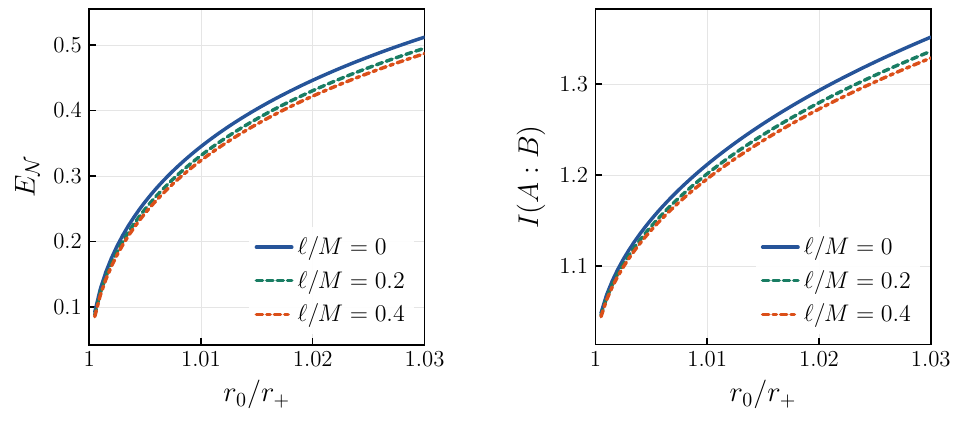}\caption{Scalar logarithmic negativity and mutual information near the outer horizon at fixed local frequency $M\omega_{\rm loc}=0.2$, for $\ell/M=0,0.2,0.4$. The observation radius is shown as $r_o/r_+$. The plotted interval is chosen within the local horizon region. Equal fractional offsets correspond to different proper distances as the geometry varies.}\label{fig:radius}\end{figure}

The principal quantum comparisons use $M$ as the unit of length and scan $0\leq\ell/M\leq0.4$. Every effective configuration in this range has a simple outer horizon. Figure~\ref{fig:geometry} compares its horizon and surface gravity with the values obtained from the complete density in Eq.~\eqref{eq:lorentzianmass}. The two calculations approach the same Schwarzschild limit. Their separation as the smearing scale grows quantifies the effect of the higher radial terms in Eq.~\eqref{eq:fullseries}. At $\ell/M=0.4$, the full outer radius is $r_{+,\rm full}/M=1.472569$, compared with $1.447214$ for the effective lapse. The corresponding surface gravities are $M\kappa_{\rm full}=0.224946$ and $M\kappa=0.213525$. The effective lapse underestimates the full surface gravity by $5.08\%$ at this endpoint.
Table~\ref{tab:fixedKilling} gives the scalar correlations and the reduced atomic emission at fixed $M\omega=0.05$. Increasing $\ell/M$ from zero to $0.4$ raises $E_{\mathcal N}$ from $0.650139$ to $0.709054$ bits and raises the mutual information from $1.488968$ to $1.554130$ bits. Over the same interval, the reduced emission probability decreases from $0.499938$ to $0.438555$. As the temperature decreases, tracing the interior leaves Alice and Bob closer to their initial Bell state, and the atom creates fewer quanta at the prescribed Killing frequency. Using the full surface gravity at $\ell/M=0.4$ gives $E_{\mathcal N}=0.689610$, $I(A:B)=1.532185$, and $\Omega^2 P_{\rm em}/g^2=0.459191$ for the same frequency and normalization. All three values lie between their Schwarzschild and effective values. At this point the exterior truncation overestimates the size of each change and preserves its direction.
\begin{table}[htbp]\centering\small\setlength{\tabcolsep}{7pt}\begin{tabular}{cccccccc}\toprule $\ell/M$ & $r_+/M$ & $M\kappa$ & $E_{\mathcal N}$ & $I(A:B)$ & $F_e$ & $\Omega^2P_{\rm em}^{(0)}/g^2$ & $\mathcal C_+^{-1}$\\ \midrule 0.0 & 2.000000 & 0.250000 & 0.650139 & 1.488968 & 0.609335 & 0.499938 & 1.000000\\ 0.1 & 1.894427 & 0.249224 & 0.651299 & 1.490213 & 0.610717 & 0.498761 & 1.055728\\ 0.2 & 1.774597 & 0.245967 & 0.656210 & 1.495502 & 0.616560 & 0.493766 & 1.127017\\ 0.3 & 1.632456 & 0.237327 & 0.669573 & 1.510028 & 0.632438 & 0.480060 & 1.225148\\ 0.4 & 1.447214 & 0.213525 & 0.709054 & 1.554130 & 0.679043 & 0.438555 & 1.381966\\ \bottomrule\end{tabular}\caption{Effective geometry, scalar correlations, and reduced HBAR emission at fixed $M\omega=0.05$. Correlation entropies are measured in bits. The last column is the factor entering the fixed $\ell$ ADM energy prescription in Eq.~\eqref{eq:hbar_area_paths}.}\label{tab:fixedKilling}\end{table}
Figure~\ref{fig:protocols} compares three observations matched at the Schwarzschild point $r_o/M=2.02$, $M\omega=0.05$. Holding the areal radius fixed while the horizon recedes raises the Killing frequency selected by the same local clock. At $\ell/M=0.4$ the logarithmic negativity reaches $0.998244$ bits in this prescription. Holding the proper distance fixed at its initial value $\varrho/M=0.400666$ instead gives $0.651482$ bits, only slightly above the Schwarzschild value $0.650139$. Equation~\eqref{eq:fixedpropersign} explains both the sign and the weak initial variation of the fixed distance curve.
Figure~\ref{fig:radius} holds the fractional offset and the local frequency fixed when comparing the geometries at a point on the horizontal axis. The same $r_o/r_+$ then samples a different proper distance in each geometry. Equation~\eqref{eq:fractionaloffset} gives the minimum of the accessible correlations as a function of $\ell/M$ for each such offset. For example, at $r_o/r_+=1.01$ the minimum occurs at $\ell/M=0.371314$. Table~\ref{tab:fourprotocols} gives the resulting correlations for all four prescriptions at a common reference frequency.
\subsection{Atomic spectrum, entropy growth, and the area factor}

\begin{figure}[htbp]\centering\includegraphics[width=\textwidth]{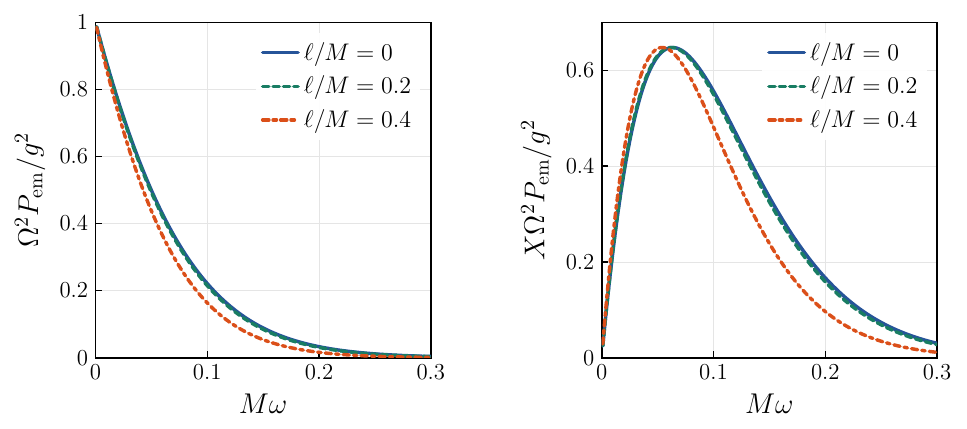}\caption{Reduced excitation and emission probability $\mathcal P_{\rm em}$ and its thermal weight $X\mathcal P_{\rm em}$ for $\ell/M=0,0.2,0.4$. The frequency is measured at infinity. The coupling and atomic gap have been divided out. The maximum of the second weight follows Eq.~\eqref{eq:thermalweightpeak}. The complete field current follows Eq.~\eqref{eq:hbar_populations}.}\label{fig:spectrum}\end{figure}

\begin{figure}[htbp]\centering\includegraphics[width=\textwidth]{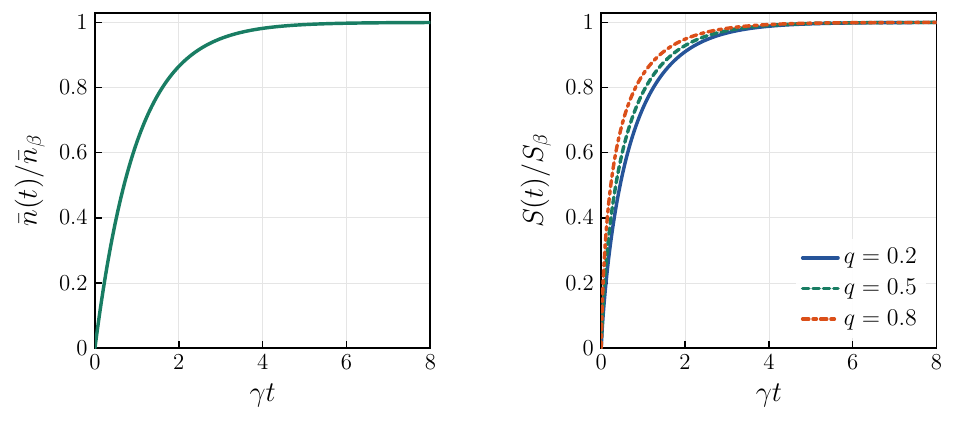}\caption{Occupation and entropy during relaxation from the exterior packet vacuum. The left plot shows the universal fraction $\bar n(t)/\bar n_\beta=1-\e^{-\gamma t}$. The right plot shows $S_\omega(t)/S_{\omega,\beta}$ for $q=0.2,0.5,0.8$. Entropy uses natural logarithms, and the horizontal variable is $\gamma t$, with $t$ measured at infinity.}\label{fig:relaxation}\end{figure}

\begin{figure}[htbp]\centering\includegraphics[width=\textwidth]{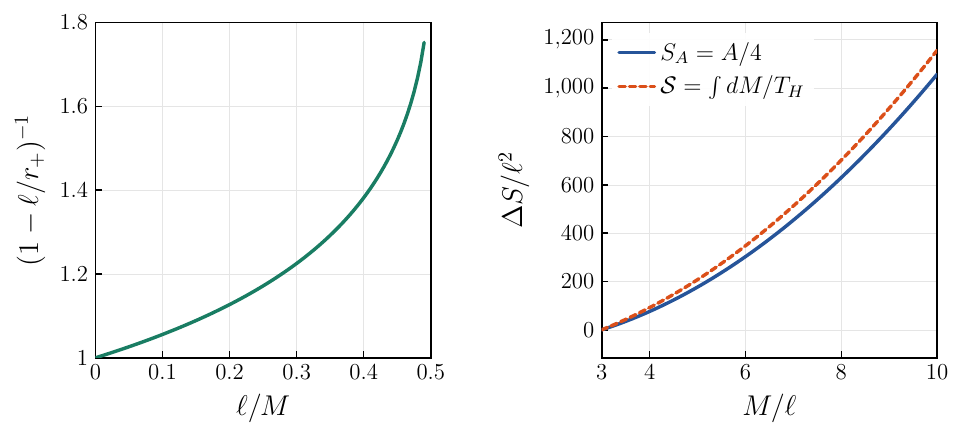}\caption{Source response and entropy increments in the effective Lorentzian family. The first plot shows $\mathcal C_+^{-1}=(1-\ell/r_+)^{-1}$. The second compares the area entropy $S_H=A/4$ with $\mathcal S_M=\int\dd M/T_H$ at fixed $\ell$, using the common reference $M=3\ell$. The two quantities follow Eqs.~\eqref{eq:hbar_source_factor} and \eqref{eq:hbar_mass_integral}, with their different energy prescriptions stated explicitly.}\label{fig:area}\end{figure}

For a common packet normalization and atomic gap, define the reduced emission $\mathcal P_{\rm em}(X)=\Omega^2P_{\rm em}^{(0)}/g^2=X/(\e^X-1)$, where $X=\omega/T_H$. Figure~\ref{fig:spectrum} shows this probability together with the thermal weight $\mathcal W(X)=X\mathcal P_{\rm em}(X)$. The weight attaches the incremental entropy coefficient $\beta_H\omega$ in Eq.~\eqref{eq:hbar_entropy_flux} to a single emission probability. For a mode initially in the vacuum, Eq.~\eqref{eq:hbar_entropy_closed} gives $\beta_H\dot E_\omega=(\mathcal R_{\rm at}g^2/\Omega^2)\mathcal W(X)\e^{-\gamma t}$. The plotted weight thus sets the initial rate of energy gain in units of $T_H$. The irreversible contribution to the entropy rate is given separately by $\sigma_\omega$.
\begin{equation} 
\begin{aligned} 
\mathcal W(X)&=\frac{X^2}{\e^X-1},\qquad \mathcal W'(X_{\rm peak})=0\ \Longleftrightarrow\ 2(1-\e^{-X_{\rm peak}})=X_{\rm peak},\\ X_{\rm peak}&=1.59362426004,\qquad \omega_{\rm peak}=X_{\rm peak}T_H.
\end{aligned} 
\label{eq:thermalweightpeak}
\end{equation}
The maximum in Eq.~\eqref{eq:thermalweightpeak} refers to a single packet with the stated normalization. Its frequency decreases in proportion to $T_H$. A broadband flux would also depend on the density of modes and their transmission through the exterior potential.
In Fig.~\ref{fig:relaxation}, the scaled time $\gamma t$ makes the occupation fractions coincide. In coordinate time, the parameters have a different effect. At fixed $M$, $\omega$, $g$, $\Omega$, and injection rate, increasing $\ell$ lowers the initial growth rate $\dot{\bar n}(0)=\Gamma_{\rm em}$ and lowers the equilibrium occupation. Yet $\gamma=\mathcal R_{\rm at} g^2(\omega/T_H)/\Omega^2$ increases, so the mode reaches a given fraction of its smaller equilibrium occupation sooner. These statements use the leading transition probabilities. The entropy fractions depend on $q$ even in scaled time because $S_\omega$ is nonlinear in $\bar n$.
Figure~\ref{fig:area} separates the area entropy from the mass over temperature integral along a fixed smearing family. The factor $\mathcal C_+^{-1}$ increases with $\ell/M$, as already indicated by Table~\ref{tab:fixedKilling}. In particular, its order $\sqrt{\Theta}$ correction precedes the order $\Theta$ change in the fixed frequency thermal spectrum. The second plot displays entropy increments from a common reference mass, so the arbitrary additive constant in Eq.~\eqref{eq:hbar_mass_integral} cancels. The difference follows from the energy assigned to the changing source. This interpretation connects the Lorentzian calculation with the source correction of Ref.~\cite{MaZhao2014} and supplies the appropriate comparison with HBAR entropy studies in other modified geometries \cite{Sen2022HBAR,Jana2024HBAR,OvgunPantig2026HBAR}.

\subsection{Reproducible calculation and checks}

The accompanying Mathematica calculation evaluates the horizon equation, surface gravity, proper distance, normalization of the quantum states, and source variation. The scalar sums use the error bounds in Eqs.~\eqref{eq:Ntail} and \eqref{eq:entropytail}. An independent calculation reproduces the scalar correlations, enclosed mass, proper distance derivative, and source factor.

For a specific finite interaction check, set $M\omega=0.05$, $\ell/M=0.2$, $\mathcal E=1$, $\chi=\e^{-x/L}$ with $L/M=0.02$, and $w=r_+/r$. The regular trajectory equations are integrated before evaluating the oscillatory amplitude. The substitution $x=-iy$ followed by $y=z/\Omega$ gives a damped contour representation of the same analytic packet integral. Table~\ref{tab:switched} shows its approach to the leading thermal result, $\Omega^2P_{\rm em}^{(0)}/g^2=0.493766$, $\Omega^2P_{\rm abs}^{(0)}/g^2=1.771009$, and $q=0.278805$. At $M\Omega=5000$, the leading switched Mellin expression alone gives emission and probability ratio changes of $0.3973\%$ and $0.8164\%$. The corresponding shifts computed along the full trajectory are $0.3977\%$ and $0.8177\%$, as predicted by Eqs.~\eqref{eq:hbar_switched_correction} and \eqref{eq:hbar_switched_ratio}, where $1/L$ dominates $r_-/(r_+\Delta_h)$. For this case their values are $50/M$ and $0.0820/M$, respectively. The last column samples the radial potential on the endpoint scale $x=\Omega^{-1}$ to estimate the importance of scattering over the part of the trajectory that dominates the integral. All rows use $w=r_+/r$. 

\begin{table}[htbp]\centering\small\begin{tabular}{cccccc}\toprule $M\Omega$ & $\Omega^2P_{\rm em}^{\rm sw}/g^2$ & $\Omega^2P_{\rm abs}^{\rm sw}/g^2$ & $P_{\rm em}^{\rm sw}/P_{\rm abs}^{\rm sw}$ & Emission change & $\varepsilon_{\rm opt}$\\ \midrule 500 & 0.508957 & 1.683187 & 0.302377 & $3.08\%$ & 0.10866\\ 1000 & 0.502615 & 1.730873 & 0.290383 & $1.79\%$ & 0.05444\\ 5000 & 0.495730 & 1.763630 & 0.281085 & $0.40\%$ & 0.01091\\ \bottomrule\end{tabular}\caption{Finite switched detector response for the trajectory and packet specified in the text. The reference is the gamma function limit in Eq.~\eqref{eq:hbar_single_probabilities}. The radial envelope and finite interaction are included in every row. The local optical parameter is $\varepsilon_{\rm opt}=V_0(r_++\Omega^{-1})/\omega^2$.}\label{tab:switched}\end{table}


\section{Conclusion}

We studied entanglement redistribution and HBAR entropy in the effective exterior geometry obtained from Lorentzian non--commutative smearing. At fixed mass, the outer radius shifted at order $\sqrt{\Theta}$, while the surface gravity first changed at order $\Theta$. Comparison with the complete Lorentzian density quantified the exterior approximation and reproduced the leading thermal correction.

Tracing over the interior partner gave a bosonic amplifier channel. Alice remained maximally entangled with the combined exterior and interior system, while the mutual information shared with each partner obeyed $I(A:B)+I(A:C)=2$ bits. At fixed mass $M$ and Killing frequency $\omega$, increasing $\ell$ lowered the thermal parameter $q=e^{-2\pi\omega/\kappa}$ and enhanced all three measures. For $M\omega=0.05$, increasing $\ell/M$ from zero to $0.4$ raised the logarithmic negativity from $0.650139$ to $0.709054$ bits and the mutual information from $1.488968$ to $1.554130$ bits. The specified fermionic occupation encoding exhibited the same trend and a finite logarithmic negativity in the limit $q\to1$.

The observer comparisons showed how the position prescription affected the first measurable change in the ideal channel. At fixed local frequency and areal radius, the redshift produced a term of order $\sqrt{\Theta}$. At fixed proper distance, the linear term in $\ell$ canceled at every finite distance, and the leading near horizon result depended only on $\omega_{\rm loc}\varrho$. At fixed fractional horizon offset, the competition between the horizon displacement and temperature produced a minimum of the accessible correlations.

For the freely falling atom, the logarithmic outgoing phase gave the same horizon exponential in the emission to absorption ratio. At fixed Killing frequency and atomic gap, increasing $\ell$ suppressed the leading emission probability, which was independent of the atom's initial specific energy. A finite interaction also sampled the radial profile of the selected mode. For the exponentially switched spherical profile, the first finite gap correction accounted for the observed approach to detailed balance. Random atomic injection drove the selected oscillator from its vacuum to a geometric stationary distribution. Its irreversible entropy production was quadratic in the occupation perturbation near equilibrium, yielding $\mathrm dS_{\rm HBAR}\simeq\mathrm dE_{\rm rad}/T_H$ to first order.

Finally, the mass dependence of the effective source supplied the factor $\mathcal{C}_+=1-\ell/r_+$ in the fixed $\ell$ identity $T_H\,\dd(A/4)=\mathcal{C}_+\dd M$. Assigning the outward radiative energy to horizon heat recovered $\dd S_{\mathrm{HBAR}}\simeq-\dd A_H^{\mathrm{rad}}/4$, while assigning it to the ADM mass loss gave $\dd S_{\mathrm{HBAR}}\simeq-\dd A_H^{\mathrm{rad}}/(4\mathcal{C}_+)$. For a finite area loss, integrating the varying coefficient yielded a logarithm of $(r_i-\ell)/(r_f-\ell)$ in the radiative entropy. The source factor introduced a correction of order $\sqrt{\Theta}$ into the entropy per unit area lost, while the fixed frequency thermal response began at order $\Theta$.

\appendix
\section{A specified fermionic occupation encoding}\label{app:fermion}

The scalar calculation can be compared with one outgoing fermionic particle oscillator and its interior antiparticle partner. We fix the occupation ordering as $(A,B,C)$, where Alice holds the reference mode, Bob accesses the exterior particle, and $C$ labels its interior partner. The Bell resource has even total fermionic parity. The quantities below are the qubit information measures of this ordered occupation representation, following the mode constructions of Refs.~\cite{Alsing2006,Bruschi2010}. This specification makes the subsystem choice explicit and permits a direct comparison with the scalar resource.

For the same positive Killing frequency and surface gravity as in Eq.~\eqref{eq:bogoliubov}, the selected pair has the isometry
\begin{equation} 
\ket0_H=c\ket{0,0}_{BC}+s\ket{1,1}_{BC},\qquad \ket1_H=\ket{1,0}_{BC},\qquad c^2=\frac1{1+q},\qquad s^2=\frac q{1+q}. 
\label{eq:fermionisometry}
\end{equation}
Here, $c$ and $s$ are real and positive. Their squared sum equals unity, and the exterior occupation of the regular vacuum is $s^2$. Applying Eq.~\eqref{eq:fermionisometry} to the Bell resource and tracing $C$ gives, in the basis $\{\ket{00},\ket{01},\ket{10},\ket{11}\}$,
\begin{equation} 
\rho_{AB}^{(F)} = \frac12\begin{pmatrix}c^2&0&0&c\\0&s^2&0&0\\0&0&0&0\\c&0&0&1\end{pmatrix},\qquad K_0=c\ket0\bra0+\ket1\bra1,\qquad K_1=s\ket1\bra0. 
\label{eq:fermiondensity}
\end{equation}
The operators in Eq.~\eqref{eq:fermiondensity} satisfy $K_0^\dagger K_0+K_1^\dagger K_1=\mathbb I_2$. The channel transfers part of the initial vacuum occupation into the exterior one particle state and multiplies its off diagonal coherence by $c$.

The joint spectrum contains $(1+c^2)/2$, $s^2/2$, and two zero eigenvalues. The partial transpose has eigenvalues $1/2$, $1/2$, $c^2/2$, and $-c^2/2$. These values give
\begin{equation} 
\mathcal N_F=\frac1{2(1+q)},\qquad E_{\mathcal N,F}=\log_2\!\left(\frac{2+q}{1+q}\right),\qquad F_{e,F}=\frac{[1+(1+q)^{-1/2}]^2}{4}.
\label{eq:fermionmeasures}
\end{equation}
Writing the binary entropy as $h_2(x)=-x\log_2x-(1-x)\log_2(1-x)$, the marginal and joint spectra also give
\begin{equation}
I_F(A:B)=1+h_2\!\left(\frac{1}{2(1+q)}\right)-h_2\!\left(\frac{q}{2(1+q)}\right),\qquad S_F(A|B)=1-I_F(A:B). 
\label{eq:fermionMI}
\end{equation}
Equations~\eqref{eq:fermionmeasures} and \eqref{eq:fermionMI} show that, at fixed $M$ and $\omega$, increasing $\ell$ lowers $q$ through Eq.~\eqref{eq:kappamonotone} and raises the negativity, mutual information, and Bell fidelity.

\begin{figure}[htbp]\centering\includegraphics[width=\textwidth]{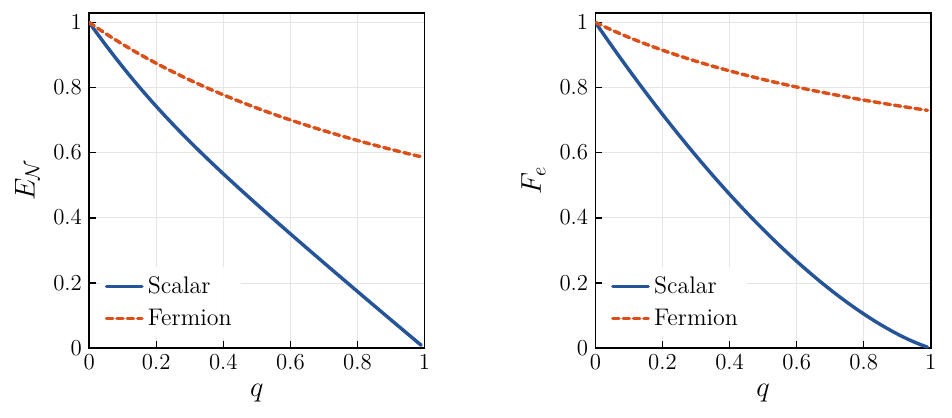}\caption{Logarithmic negativity and Bell entanglement fidelity for the scalar resource and the specified fermionic occupation encoding. Both use the same thermal parameter $q$. The scalar and fermionic formulas follow Eqs.~\eqref{eq:negativity}, \eqref{eq:fidelity}, and \eqref{eq:fermionmeasures}, with their subsystem choices stated explicitly.}\label{fig:statistics}\end{figure}

Both statistics recover the Bell values as $q\to0$. As $q\to1$, this fermionic encoding gives $\mathcal N_F\to1/4$, $E_{\mathcal N,F}\to\log_2(3/2)$, $I_F\to1$, and $F_{e,F}\to(3+2\sqrt{2})/8$. Both exterior resources carry one bit of mutual information in this limit. The scalar has $E_{\mathcal N}\to0$ and $F_e\to0$, whereas the specified fermionic pair preserves finite values of both quantities. The comparison holds for the specified occupation encoding and mode ordering. Including further spin or antiparticle modes would define a different resource.
Figure~\ref{fig:statistics} displays the two statistics at the same thermal parameter.



\preparepaperbibliography
\bibliography{main}

\end{document}